\documentstyle [12pt]{article}
\begin{document}
\vspace{7mm}

\begin{center}

{\Large\bf  Quantum Black Hole Model and Hawking's Radiation }

\vspace{12mm}
{\large Victor Berezin }
\vspace{3mm}

Institute for Nuclear Research of the Russian Academy of Sciences,

60th October Anniversary Prospect, 7a, 117312, Moscow, Russia

e-mail: berezin@ms2.inr.ac.ru
\end{center}

\vspace{14mm}
\begin{center}
{\large\bf  Abstract.}
\end{center}
\vskip0.5cm
The black hole model with a self-gravitating charged spherical symmetric dust thin shell as 
a source is considered. The Schroedinger-type equation for such a model is derived. This 
equation appeared to be a finite differences equation. A theory of such an equation is 
developed and general solution is found and investigated in details. The discrete spectrum 
of the bound state energy levels is obtained. All the eigenvalues appeared to be infinitely 
degenerate. The ground state wave functions are evaluated explicitly. The quantum black 
hole states are selected and investigated. It is shown that the obtained black hole mass 
spectrum is compatible with the existence of Hawking's radiation in the limit of low 
temperatures both for large and nearly extreme Reissner-Nordstrom black holes. The 
above mentioned infinite degeneracy of the mass (energy) eigenvalues may appeared helpful 
in resolving the well known information paradox in the black hole physics.  
\newpage
{\large\bf I.Introduction.}
\vskip1cm
The fate of black holes became a subject of interest since Hawking's discovery 
\cite{Hawking} that the large black holes should emit a blackbody radiation with the 
temperature depending on their parameters. The problem is that the temperature of the 
(uncharged, nonrotating) black hole grows when its mass decreases, and the density of the 
emitting radiation can no longer be considered as negligible, thus requiring an account of 
the back reaction of the radiation on the black hole metric. To treat such a 
self-consistent problem is an extremely difficult task.

Moreover, the values of densities and sizes of the objects involved are such that the 
classical gravity needs quantum gravity corrections.

There is a general belief that the back reaction and/or the quantum gravity corrections 
would  prevent the black holes from complete evaporation. This would lead also to 
numerous cosmological consequences.

The full treatment requires the field theoretical or (super) string consideration.

In the present paper we try a quite different approach. We consider a spherically 
symmetric black hole (both charged and uncharged) as a quantum mechanical object (like a 
hydrogen atom). We made the simplest possible choice for its source - a self-gravitating 
spherically symmetric dust shell. The thin shell limit in General Relativity elaborated in 
a geometrically invariant form by W.Israel \cite{Werner} make many problems exactly 
solvable and, though very simple, allows one to obtain important physical results both in 
black hole physics \cite{BHp} and cosmology \cite{cosmol}.

The thin shell model allows us to reduce the problem in the spherically symmetric case to 
the essentially the one dimensional problem which is rather easy to quantize. The 
quantization procedure used here was first proposed in the paper by V.A.Berezin, 
N.G.Kozimirov, V.A.Kuzmin and I.I.Tkachev \cite{first} and then in \cite{me}. The discrete 
mass spectrum for the self-gravitating dust shells was suggested in \cite{first}. In a paper
 \cite{Bures} it was generalized to electrically charged shells and used then to obtain 
a black hole spectrum, which appeared compatible with a Hawking's blackbody radiation in 
quasi-classical quasiclassical regime, i.e. for large as well as nearly extreme Reissner-Nordstrom black 
holes.

In the present paper we summarize the results, obtained earlier and construct for them a 
solid base. Namely, we describe a method of solving our Schroedinger equation in finite 
differences with Coulomb potential, found the general solution, solved the boundary 
conditions and proved the existence of discrete spectrum for bound states, evaluate 
explicitly the ground state wave function and showed that all the mass (energy) 
eigenvalues are infinitely degenerate in the frozen formalism.
 
The plan of the paper is the following. In Sect.II we present the classical equations of 
motion of a self-gravitating charged dust shells in General Relativity and discuss some 
of their properties. This model is quantize in Sect.III. The resulting Schroedinger equation 
is not a differential equation but the equation in finite differences. We found the 
asymptotics of its solutions in singular points (near the origin and at the infinity) in 
Sect.IV. In Sect.V we present the general solution to the equation in the momentum 
representation and discuss its properties. A nontrivial transition to the coordinate 
representation is given in Sect.VI and the fundamental solution is introduced. This solution 
is investigated in details in Sect.VII. In Sect.VII the appropriate boundary conditions at 
the origin and infinity are written to ensure the Hamiltonian be self adjoint operator on 
the positive semi-axis. In Sect.IX the discrete spectrum for bound states is found, the 
new polynomials are introduced and investigated (in particular, the generating function for 
them is evaluated). The ground state wave function is presented in Sect.X. In Sect.XI the 
notion of quantum black holes is introduced and the black hole mass spectrum is derived. 
This spectrum is shown to be compatible with the existence of Hawking's radiation in 
Sect.XII. The Paper concludes with Discussions.

Throughout the paper the system of units in which $c=\hbar =k=1$ is used, where $c$ is a 
velocity of light, $\hbar$ is a Planck's constant, $k$ is a Boltzmann's constant.

\newpage
\noindent {\large\bf II.Classical Equations of Motion for a Self-gravitating Charged
Spherically Symmetric Dust Shell.}
\vskip1cm
We start with a description of the model. Thus is just a self-gravitating spherically symmetric
dust thin shell, endowed with a bare mass $M$ and electric charge $e$. The whole spacetime is
divided into three different regions: the inner part, the outer part and the 3-dimensional 
timelike hypersurface separating them.

The general metric of a spherically symmetric spacetime has a form
\begin{eqnarray}
ds^2 = Adt^2 + 2Hdtdq + Bdq^2 r^2(t,q)d\Omega^2, \\
A \ge 0 , B \le 0 , \nonumber
\label{inter}
\end{eqnarray}

where $t$ and $q$ are correspondingly timelike and spacelike coordinates, $A$, $H$ and $B$
 are functions of $t$ and $q$ only, $r(t,q)$ is the radius of a two-dimensional sphere
 (in the sense that the area of the sphere is $4\pi r^2$0,
\begin{equation}
d\Omega^2 = d\theta^2 + \sin^2\theta d\varphi^2
\end{equation}
being the line interval of the unit sphere. 

Contrast to the flat spacetime, the normal vector to the surface $r = const$ may be not 
only spacelike  but also timelike. In the first case
\begin{equation}
\Delta \equiv g^{\alpha\beta}r_{,\alpha}r_{,\beta} < 0,
\label{norma}
\end{equation}
and the corresponding region is called $R$-region (here $g_{\alpha\beta}$ is a metric 
tensor, $g^{\alpha\beta}$ is its inverse, $r_{,\alpha}$ denotes the partial derivative
with respect to the corresponding coordinates, Greek indices run from 0 to 3). In the
flat case the $R$-region occupies the whole spacetime. In the timelike case
\begin{equation}
\Delta > 0.
\label{T}
\end{equation}
Such a region is called the $T$-region (the notions of $R$- and $T$-regions were 
introduced in \cite{Novikov64}). It easy to show that the condition $\dot r = 0$ ("dot"
denotes the time derivative) cannot be satisfied in a $T$-region, hence it should be 
either $\dot r > 0$ (this region of inevitable expansion is called $T_+$-region), or 
$\dot r < 0$ (inevitable contraction, a $T_-$-region). Correspondingly, it is impossible 
to get $r^\prime = 0$ ("prime" denotes the spatial derivative) in $R$-regions, and a region
 with $r^\prime > 0$ is called an $R_-$-region, while that with $r^\prime < 0$ is an $R_-$-
region. The $R_+$ and $R_-$ regions correspond to different sides of an Einstein-Rosen 
bridge (see, e.g., \cite{MTW} and references cited therein).

The solution to  Einstein equations representing the Reissner-Nordstrom (spherically
symmetric, charged) black hole is well known and can be put in the form
\begin{equation}
ds^2 = f dt^2 - f^{-1} dr^2 - r^2 d\Omega^2,
\label{R-N}
\end{equation}
where
\begin{equation}
f = 1 - \frac{2\kappa m}{r} + \frac{\kappa e^2}{r^2},
\label{f}
\end{equation}
and $m$ is the total mass (energy) of the system, {e} is its electrical charge, $\kappa$ 
is the gravitational constant (equal to the inverse square of Planckian mass,  
$\kappa = M^{-2}_{\it pl}$, in the chosen units; note also that in these units the 
electric charge is dimensionless, and radius has dimension of inverse mass). For the zero 
charge case the above metric is reduced to the Schwarzschild solution. The
Reissner-Nordstrom metric has been extensively studied (see \cite{MTW} for the detailed 
description and references). What is important for us is the following.

The invariant $\Delta$, introduced above, Eqn.(\ref{norma}), is
\begin{equation}
\Delta = -f = -1 + \frac{2\kappa m}{r} - \frac{\kappa e^2}{r^2}.
\label{delta}
\end{equation}

This function has two different positive roots, if $\sqrt{\kappa}m > |e|$,
\begin{eqnarray}
r_+ = \kappa m + \sqrt{\kappa^2 m^2 - \kappa e^2}, \nonumber\\
\nonumber\\
r_- = \kappa m - \sqrt{\kappa^2 m^2 - \kappa e^2}. 
\label{roots}
\end{eqnarray}

The surfaces $r_{\pm} = const$ are null surfaces separating regions with $\Delta < 0$ 
($R$-regions, $r > r_+, 0 < r_-$ ) from that with $\Delta > 0$ ($T$-regions, $r_- < r < r_+$)
 The $r_+ = const$ surface is the event horizon, the so-called outer horizon, like the event 
horizon in the Schwarzschild solution to which it reduces when $e = 0$. The inner horizon,
$r_- = const$, was recognized as a Cauchy horizon beyond which the solution cannot be 
continued unambiguously. There are infinitely many $R_\pm$ and $T_\pm$ regions. They form 
some kind of ladder on the Carter-Penrose diagram for a maximally analytically extended 
Reissner-Nordstrom spacetime. But due to the generic instability of the Cauchy horizons 
only one portion (in between two nearest Cauchy horizons) can be considered as physical 
(for the details and references see \cite{MTW} and \cite{NovFr} ). In many aspects this part 
is the same as the maximal analytical extension of the Schwarzschild metric. The main point 
for us is that there are two asymptotically flat $R$-regions (where $r > r_+$), one of them, 
$R_+$, is the region external to the black hole, it is "our" side of the Einstein-Rosen 
bridge, the "other" side being $R_-$-region and is called also a wormhole region \cite {MTW}.

The Reissner-Nordstrom black hole with $\sqrt{\kappa}m = |e|$ is called the extreme black 
hole. For $\sqrt{\kappa}m < |e|$ no black hole solution exists.

The metric of the 3-dimensional surface, $\Sigma$, representing the evolution of the thin 
shell can be written as
  
\begin{equation}
ds^2_{|\Sigma} = d\tau^2 - \rho^2(\tau)d\Omega^2,
\label{Sigma}
\end{equation}
where $\tau$ is the proper time measured by an observer comoving to the shell. This surface 
may be considered as embedded into 4-dimensional spacetime, then the Einstein equations on 
the shell may be expressed in terms of the extrinsic curvature tensor $K_{ij}$, the 
curvature of embedding (Latin indices take values 0,2 and 3), and surface energy tensor 
$S_{ij}$ \cite{Werner}:
\begin{equation}
[K^j_i] - \delta^j_i [K^l_l] = - 8\pi\kappa S^j_i,
\label{K}
\end{equation}
where
\begin{equation}
[K^j_i] = K^j_i(out) - K^j_i(in) \nonumber
\end{equation}
is the jump in the extrinsic curvature across the shell (the coordinate, $n$, normal 
to the surface is chosen increasing in the outward direction). The remaining Einstein
equations, $\left(i\atop n\right)$ and $\left(n\atop\ n\right)$ components, may be 
written in the form (vertical bar denotes covariant differentiation with respect to 
the metric on $\Sigma$):
\begin{equation}
S^j_{i|j} + [T^n_i] = 0,
\label{i-n}
\end{equation}
\begin{eqnarray}
{K^i_j} S^j_i + [T^n_n] = 0,\\  
{K^i_j} = {\frac{1}{2}}(K^i_j(out) + K^i_j(in))
\label{n-n}
\end{eqnarray}
and coincide with the covariant energy-momentum conservation equation written for the 
shell \cite{bubble}. In the case of the spherical symmetry Eqns.(\ref{K}-\ref{n-n}) can 
be expressed in terms of invariants and take the form \cite{bubble} (note that now 
$A^2_2 = A^3_3$ and $A^0_0 = inv$ for any 3-tensor on $\Sigma$ ):

\begin{equation}
[K^2_2] = 4\pi\kappa S^0_0 ,
\label{S0}
\end{equation}

\begin{equation}
[K^0_0] + [K^2_2] = 8\pi\kappa S^2_2 ,
\label{S2}
\end{equation} 

\begin{equation}
\dot S^0_0 + 2 \frac{\dot \rho}{\rho} (S^0_0 - S^2_2) + [T^n_0] = 0 ,
\label{Sdot}
\end{equation} 
 where
\begin{equation}
K^0_0 = - \frac{\sigma}{\sqrt{\dot \rho^2 - \Delta}} 
\left( \ddot \rho + \frac{1 + \Delta}{2\rho} - 4\pi\kappa\rho T^n_n \right) ,
\label{K0}
\end{equation} 

\begin{equation}
K^2_2 = - \frac{\sigma}{\rho} \sqrt{\dot \rho^2 - \Delta} .
\label{K2}
\end{equation}
Here "dot" denotes the derivative with respect to the proper time $\tau$,  $\sigma = \pm 1$
 (its meaning will be explained in a moment), and $T^n_0$ and $T^n_n$ are normal flow and
normal stress components of the energy-momentum tensor external (internal) to the shell. 
For the Coulomb field
\begin{equation}
T^0_0 = T^1_1 = \frac{e^2}{8\pi r^2} ,\qquad T^1_0 = 0
\label{Coulomb}
\end{equation}  
for any choice of the (spherical) coordinate system. For a dust shell $S^2_2 = 0$, by
definition. So,
\begin{equation}
S^0_0 = \frac{M}{4\pi\rho^2(\tau)} ,
\label{dustS}
\end{equation} 
where $M$ is identified as the bare mass of the shell. As can be easily shown, 
Eqn(\ref{S2}) is a differential consequence of the Eqns.(\ref{S0}) and (\ref{Sdot}),
and we get eventually the single equation (inserting for $\Delta$ Eqn.(\ref{delta}) and
remembering that inside the shell the spacetime is empty)
\begin{equation}
\sigma_{in}\sqrt{\dot \rho^2 + 1} - \sigma_{out}\sqrt{\dot \rho^2 + 1 - 
\frac{2\kappa m}{\rho} + \frac{\kappa e^2}{\rho^2}} = \frac{\kappa M}{\rho}
\label{shelleq}
\end{equation}
which is the main subject of our future consideration.

The quantity $\sigma$ appeared for the first time in the expression for extrinsic 
curvature tensor, Eqns.(\ref{K0}) and (\ref{K2}). Its meaning is the following.
 $\sigma = +1$ if radii increase in the outward normal direction to the shell, and 
$\sigma = -1$ if radii decrease. Therefore, in the $R$-regions $\sigma$ does not
 change its sign, the latter being the sign of an $R$-region. It is clear now that
 on "our" side of the Einstein-Rosen bridge we have $\sigma = +1$, this we shall 
call the "black hole case", while on the "other" side $\sigma = -1$, this we shall 
call the "wormhole case". In the flat(Minkowski" spacetime we have everywhere 
$\sigma = +1$.

Solving Eqn.(\ref{shelleq}) for $m$ we get
\begin{equation}
m = M\sqrt{\dot \rho^2 + 1} - \frac{\kappa M^2 - e^2}{2\rho} ,
\label{mass}
\end{equation} 
and now it is seen that this is nothing more but the energy conservation equation, 
the square root being the Lorentz factor written in terms of proper time derivatives.

We shall be interested in the bound motion only. Let $\rho_0$ denotes the value of 
radius of the shell at the moment of rest and $\rho_0 > r_+$ (i.e. we are outside 
the event horizon). Then it follows from Eqns.(\ref{shelleq}) and (\ref{mass}) that
\begin{equation}
m = M - \frac{\kappa M^2 - e^2}{2\rho_0} ,
\label{m}
\end{equation}

\begin{equation}
\sigma_{out} = sign \left(1 - \frac{\kappa M}{\rho_0}\right) ,
\label{sigmaout}
\end{equation}

\begin{equation}
\frac{\partial m}{\partial M} = 1 - \frac{\kappa M}{\rho_0} .
\label{mprime}
\end{equation}
It is seen now that the total mass $m$ increases with bare mass $M$ when 
$\sigma _{out} = +1$, i.e. in the black hole case, and it increases when 
$\sigma_{out} = -1$ (the wormhole case). In the black hole case

\begin{equation}
\frac{m}{M} > \frac{1}{2} + \frac{e^2}{2\kappa M^2} 
\label{m/M}
\end{equation}
thus ranging from 1/2 ($M \to \infty$ ) to 1 ($M \to e/\sqrt{\kappa}$, and 
$m \to e/\sqrt{\kappa}$ as well ). The opposite inequality is valid in the wormhole 
case.

To complete our discussion let us consider the thin shell with negative total mass, 
$m < o$  (but positive bare mass $M$), From Eqn.(\ref{shelleq}) it is readily seen that 
$\sigma_{out} = -1$ everywhere, and the spacetime topology is $S^3 \times R^1$ rather
than $R^3 \times R^1$ as it is for the positive mass $m$. Due to different topologies 
the cases $m > 0$ and $m < 0$ in quantum theory should be treated separately. In what 
follows we will consider only positive masses, $m < 0$.
\newpage
{\large\bf III. Quantization.}
\vskip1cm

To quantize a system we have to construct first the classical Hamiltonian and then to 
replace the classical Poisson brackets by the quantum commutation relations. In making 
all these we will follow the procedure elaborated in \cite{first} and \cite{me}. 

Since $m$ is a total mass of the system and Eqn.(\ref{mass}) represents the energy 
conservation law it is reasonable to consider it numerically equal to the Hamiltonian 
of the system in question. Then the conjugate momentum $p$, Lagrangian $L$ and 
Hamiltonian $H$ can be found from the following relations
\begin{equation}
p = \frac{\partial L}{\partial {\dot \rho}} ,
\label{moment}
\end{equation}

\begin{equation}
H = p\dot \rho - L = \dot \rho\frac{\partial L}{\partial {\dot \rho}} - L ,
\label{Hamilt1}
\end{equation}

\begin{equation}
L = \dot \rho \int H \frac{d{\dot \rho}}{{\dot \rho}^2} = 
\dot \rho \int \frac{\partial H}{\partial {\dot \rho}} \frac{d{\dot \rho}}{\dot \rho} - H
\label{Lagran}
\end{equation}
which give for the conjugate momentum
\begin{equation}
p = \int \frac{\partial H}{\partial{\dot \rho}} \frac{d {\dot \rho}}{\dot \rho} .
\label{pif}
\end{equation}
Substituting for $H$ the Eqn.(\ref{mass}) one gets
\begin{equation}
p = M \log (\dot \rho + \sqrt{{\dot \rho}^2 + 1}) + F(\rho) ,
\label{pi}
\end{equation}

\begin{equation}
L = M (\dot \rho \log(\dot \rho + \sqrt{{\dot \rho}^2 + 1}) - \sqrt{{\dot \rho}^2 + 1}) + 
{\dot \rho} F(\rho) ,
\label{L}
\end{equation}
where $F(\rho)$ is an arbitrary function. The choice of this function does not affect
the Lagrangian equations of motion, and leads to the canonically equivalent systems for
 Hamiltonian approach. This choice may become important when one explores the rather  
complex structure of maximal analytical extensions of black hole spacetimes trying to 
construct a wave function covering the whole manifold as was shown for a specific example 
in \cite{me}. But here we do not need such a complication and $F(\rho) = 0$. 
From Eqn.(\ref{pi}) we get for $\dot \rho$

\begin{equation}
\dot \rho = \sinh\frac{p}{M} ,
\label{rdot}
\end{equation}
and after inserting this into Eqn.(\ref{mass}), the Hamiltonian $H$ is

\begin{equation}
H = M \cosh\frac{p}{M} - \frac{\kappa M^2 - e^2}{2\rho} .
\label{H}
\end{equation}
It is convenient to make a canonical transformation to the new variables ($M$ is the
bare mass of the shell)

\begin{equation}
x = M \rho , \qquad \Pi = \frac{1}{M} \rho ,
\label{xPi}
\end{equation}
then
\begin{equation}
H = M \left( \cosh\Pi - \frac{\kappa M^2 - e^2}{2x}\right) .
\label{newH}
\end{equation}

Let us now impose the quantum commutation relation

\begin{equation}
[\Pi, x ] = -i .
\label{commut}
\end{equation}
Then, using the coordinate representation with $\Pi = - i\partial/\partial x$ and the
 relation

\begin{equation}
e^{-i\frac{\partial}{\partial x}} \Psi(x) = \Psi(x-i)
\label{shift}
\end{equation}
we obtain the Schroedinger stationary equation

\begin{equation}
H\Psi(x) = E\Psi(x),
\label{quantum}
\end{equation}

\begin{equation}
\frac{M}{2}\left( (\Psi(x+i) + \Psi(x-i) - \frac{\kappa M^2 - e^2}{x} \Psi(x)\right) = 
m\Psi(x) .
\label{mine}
\end{equation}
Introducing the notation $\epsilon = m/M$ we can rewrite this as 

\begin{equation}
\Psi(x+i) + \Psi(x-i) = \left(2\epsilon + \frac{\kappa M^2 - e^2}{x}\right) \Psi(x) .
\label{mefinal}
\end{equation}
Some remarks are in order. First, the Schroedinger equation obtained is not the differential,
but the equation in finite differences. This is a direct consequence of the quantization 
in proper time which is quite natural in the framework of General Relativity. Second, the
total mass $m$ is not arbitrary anymore but it is an eigenvalue of Hamiltonian operator, 
subject to the condition that for corresponding eigenfunctions the Hamiltonian is Hermitian 
on the positive semi-axes. Third, the Hamiltonian picture we got may not be equivalent to 
that which could be obtained directly from the Einstein-Hilbert action because of making 
use of the proper time definition of which depends on the history of the shell. It would 
definitely inequivalent if we would start from the Einstein-Hilbert action and express it 
in terms of proper time before making variations. But we made it the other way around. 
We started from equations of motion written already in proper time and constructed the 
new Lagrangian which results in just the equations in proper time with no restrictions 
or/and constraints. The procedure is thus self-consistent. We believe then that the resulting 
mass spectrum will be correct because it depends on the existence or nonexistence of the 
appropriate solutions but not on their detailed behavior.

It should be noted also that the step in our finite difference equation is along the 
imaginary axes, so the solutions should have certain analyticity to be continued to the 
complex plane.

With all dimensional quantities restored the shifted argument ($x\pm i$) becomes
\begin{equation}
x \pm i \to M \left (\rho \pm \frac{1}{M}i\right) \to \rho \pm \frac{\hbar}{Mc}i ,\nonumber
 \end{equation}
so for $\rho \gg \hbar/Mc$ we obtain the nonrelativistic limit. Expanding Eqn.(\ref{mefinal})
 up to the second order in ($\hbar /Mc$) we get
\begin{equation}
- \frac{1}{2M}\frac{d^2 \Psi}{d\rho^2} - \frac{\kappa M^2 - e^2}{2\rho} \Psi = (E - M)\Psi
\label{nonrel}
\end{equation}
which is just the nonrelativistic Schroedinger equation for radial wave function in s-wave, 
($E-M$) being the nonrelativistic energy of the system. For negative values of ($E-M$) 
acceptable solutions exist only for a discrete spectrum of energies, namely
\begin{equation}
(E- M)_n = - \frac{M(\kappa M^2 - e^2)^2}{8n^2} ,\qquad 
n = 1,2,...
\label{Rydberg}
\end{equation}
For $\kappa = 0$ it reduces to the well known Rydberg formula for the hydrogen atom.
\newpage
{\large \bf IV. Asymptotics.}
\vskip1cm
In this Section we will find the asymptotics for the solution of the finite differences 
Schroedinger equation in coordinate representation. i.e. for the Eqn.(\ref{mefinal}).

Our strategy will be the following. Expand, first, the left had side of Eqn.(\ref{mefinal}) 
into series in derivatives. We get
\begin{eqnarray}
\Psi(x+i) = \Psi(x) + i\Psi^{\prime} + \frac{i^2}{2!}\Psi^{\prime \prime} +
\frac{i^3}{3!}\Psi^{(3)} + \cdots\\
\nonumber \\
\Psi(x-i) = \Psi(x) - i\Psi^{\prime} + \frac{i^2}{2!}\Psi^{\prime \prime} -
\frac{i^3}{3!}\Psi^{(3)} + \cdots\\
\nonumber \\
\sum_{K=1}^{\infty}\frac{(-1^K}{(2K)!}\Psi^{(2K)} = \left(\epsilon - 1 + 
\frac{\kappa M^2 - e^2}{2x}\right)\Psi(x) .   
\label{series} 
\end{eqnarray}
We truncate this series to obtain
\begin{equation}
\sum_{K=1}^N\frac{(-1^K}{(2K)!}\Psi^{(2K)} = \left(\epsilon - 1 + 
\frac{\alpha}{2x}\right)\Psi(x) .   
\label{trunc}
\end{equation}

\begin{equation}
\Psi^{(2N)} = - \sum_{K=1}^{N-1}(-1)^{N-K}\frac{(2N)!}{(2K)!}\Psi^{(2K)} + 
(-1)^N(2N)!\left(\epsilon - 1 + \frac{\alpha}{2x}\right)\Psi(x) ,
\label{psi2n}
\end{equation}
with $N$ arbitrary and $\alpha = \kappa M^2 - e^2$ .

The Eqn.(\ref{psi2n}) will be our starting point. To find the asymptotic expansion of the 
solution to this equation we shall use the matrix method described in the book \cite{Wasow}. 
First of all, we should transform Eqn.(\ref{psi2n}) into a matrix form. This can be made by 
redefinition
\begin{equation}
\begin{array}{c}
\Psi = y_1 ,\qquad  \Psi^{\prime} = y_2 ,\qquad...,\qquad \Psi^{(2N)} = y_{2N} ,\nonumber\\
\nonumber\\
\Psi^{(2N)} = y_{2N}^{\prime} = - \sum_{K=1}^{N-1}(-1)^{N-K}\frac{(2N)!}{(2K)!} y_{2K+1} + 
(-1)^N(2N)!\left(\epsilon - 1 + \frac{\alpha}{2x}\right)y_1 .
\end{array}
\label{igrec}
\end{equation}

Forming from ${y_i}$ the one-row matrix we get
\begin{equation}
Y^{\prime} = A(x)Y ,\qquad  A(x) = A_0 + \frac{1}{x}A_1 ,
\label{matrixeqn}
\end{equation}
where  
\begin{eqnarray}
Y &=& 
\left(
\begin{array}{c}
y_1\\
y_2\\
\vdots\\
y_{2N}
\end{array}
\right) ,\nonumber\\
\nonumber\\
A_0 &=& 
\left(
\begin{array}{ccccc}
0& 1& 0&\ldots & 0\\
0& 0& 1&\ldots & 0\\
\ldots&\ldots&\ldots&\ldots&\ldots\\
0& 0& 0&\ldots& 1\\
a_1&\ldots&\ldots&\ldots& a_{2N}
\end{array}
\right) ,\nonumber\\
\nonumber\\
a_1 &=& (-1)^N(2N)!(\epsilon - 1) ;\qquad  a_{2K} = 0;\nonumber\\
\nonumber\\
a_{2K+1} &=& (-1)^{N-K-1}\frac{(2N)!}{(2K)!} ;\qquad  K = 1,..., N-1 ;\nonumber\\
\nonumber\\
A_1 &=& (-1)^N(2N)!\frac{\alpha}{2} 
\left(
\begin{array}{cccc}
0& 0&\ldots& 0\\
\ldots&\ldots&\ldots&\ldots\\
0& 0&\ldots& 0\\
1& 0&\ldots&0
\end{array}
\right) .
\label{matrices}
\end{eqnarray}

The Eqn.(\ref{matrixeqn}) has two singular points. The first one, at $x = 0$, is a regular 
singular point (for all the definitions and theorem proofs see \cite{Wasow}). The second 
one, at $x = \infty$, is an irregular singular point. Their treatments need separate 
considerations.

Let us start with a regular singular point at $x = 0$. The matrix equation is usually 
presented in the following form in this case 
\begin{eqnarray}
xY^{\prime} = B(x)Y ,\\
\nonumber\\
B(x) = B_0 + xB_1 ,\\
\nonumber\\
B_0 = A_1 ;  B_1 = A_0 .
\label{xzero}
\end{eqnarray}
Because the matrix $B_0$ has only zero eigenvalues it can not be transformed to the 
diagonal form, but only to the Jordan's normal form by some similarity transformation.
The matrix $T$ corresponding to such a transformation is not unique but this does not 
affect the result. The Jordan's normal form of the transformed matrix is unique. Thus, 
with $Y = TZ$, we have
\begin{equation}
\begin{array}{c}
xZ' = CZ ;\qquad   C = C_0 + xC_1 ;\nonumber\\
\nonumber\\
C_0 = T^{-1}A_0T = H ;\nonumber\\
\nonumber\\
H  = 
\left(
\begin{array}{ccccc}
0& 1& 0&\ldots& 0\\
0& 0& 0&\ldots& 0\\
\ldots&\ldots&\ldots&\ldots&\ldots\\
0& 0& 0&\ldots&0 \\
\end{array}
\right) ;\nonumber\\
\nonumber\\
T = 
\left(
\begin{array}{ccccc}
0& 1& 0&\ldots& 0\\
0& 0& 1&\ldots& 0\\
\ldots&\ldots&\ldots&\ldots&\ldots\\
\gamma& 0& 0&\ldots& 0\\
\end{array}
\right) ;~~
T^{-1} = 
\left(
\begin{array}{ccccc}
0& 0&\ldots& 0& \gamma^{-1}\\
1& 0&\ldots& 0& 0\\
0& 1&\ldots& 0& 0\\
\ldots&\ldots&\ldots&\ldots&\ldots\\
0& 0&\ldots& 1& 0
\end{array}
\right) ;\nonumber\\
\nonumber\\
C_1 = 
\left(
\begin{array}{cccccc}
0& c_2& c_3& c_4&\ldots& c_{2N}\\
0& 0& 1& 0&\ldots& 0\\
0& 0& 0& 1&\ldots& 0\\
\ldots&\ldots&\ldots&\ldots&\ldots&\ldots\\
0& 0& 0& 0&\ldots& 1\\
\gamma& 0& 0& 0&\ldots& 0
\end{array}
\right) ,\nonumber\\
\nonumber\\
c_2 = \frac{2(\epsilon -1)}{\alpha} ; \qquad c_{2K+2} = \frac{2(-1)^{K+1}}{(2K)!\alpha} ,
\nonumber\\
\nonumber\\
\gamma = (-1)^N (2N)!\frac{\alpha}{2} .
\label{HTCmatr}
\end{array}
\end{equation}

Now the solution can be written as follows
\begin{equation}
Y = T \left( I + \sum_{r=1}^{\infty}x^rP_r\right) (I + H \log x ) ,
\label{asym0}
\end{equation}
where $I$ is the unit matrix, and the matrices $P_r$ can be calculated from the following 
recurrent equation
\begin{eqnarray}
\label{receqn}
HP_r - P_rH - rP_r = - C_1P_{r-1} ,\\
\nonumber\\
P_0 = I , \qquad  r = 1,2,....\nonumber
\end{eqnarray}

The matrix $Y$ is to be understood in that sense that each of its row gives a linearly 
independent fundamental solution to the Eqn.(\ref{matrixeqn}). It is not difficult to see 
the logarithm will appear in the first line (which is $y_1 = \Psi$ and the only one of 
interest) starting from ($1+x^{2N}\log x$)-term, and when $N \to \infty$ it disappears 
leaving $1$ (this is an unexpected and remarkable result).

The final result is that the wave function has the following asymptotics at $x \to 0$ (we 
are writing down only the leading terms)
\begin{eqnarray}
\label{xn}
\Psi = x^n\qquad\mbox{at $x \to 0$} ,\\
n = 0,1,2,....\nonumber
\end{eqnarray}

Let us now proceed with the asymptotical behavior of the solution near the irregular 
singular point $X = \infty$. The procedure will be different in this case. We start again 
from Eqn.(\ref{matrixeqn}) and make a similarity transformation. But now the matrix $A_0$ 
plays the role of $B_0(= A_1)$, and it is possible to transform it to the diagonal form 
(assuming that all its eigenvalues are different what is, fortunately, the case as will be 
clear in a moment). With $Y = TZ$ we have now
\begin{eqnarray}
Z^{\prime} = CZ ;\qquad  C = C_0 + \frac{1}{x}C_1 ,\nonumber\\
\nonumber\\
C_0 = T^{-1}A_0T = \mbox{diag}(\lambda_1,...,\lambda_{2N});\nonumber\\
\nonumber\\
T = 
\left(
\begin{array}{cccc}
1& 1&\ldots& 1\\
\lambda_1& \lambda_2&\ldots&\lambda_{2N}\\
\ldots&\ldots&\ldots&\ldots\\
\lambda_1^{2N-1}&\lambda_2^{2N-1}&\ldots&\lambda_{2N}^{2N-1}
\end{array}
\right)
\label{Ts}
\end{eqnarray}
Denoting by $\tilde t_{ij}$ the elements of the inverse matrix $T^{-1}$ we obtain for $C_1$: 
\begin{equation}
C_1 = (-1)^N(2N)!\frac{\alpha}{2}
\left(
\begin{array}{ccc}
\tilde t_{1,2N}&\ldots&\tilde t_{1,2N}\\
\ldots&\ldots&\ldots\\
\tilde t_{2N,2N}&\ldots&\tilde t_{2N,2N}
\end{array}
\right) ,
\label{C1}
\end{equation}
so the matrix $C_1$ has all the rows identical.

The characteristic equation for $\lambda$ reads as follows
\begin{equation}
\sum_{K=0}^N \frac{-1^K\lambda^{2K}}{(2K)!} = \epsilon ,
\label{lambda}
\end{equation}
which in the limit $N \to \infty$ becomes
\begin{equation}
\cos\lambda = \epsilon = \frac{m}{M} ,
\label{cos}
\end{equation}
and it is clear now that all the eigenvalues are different. Note also that for $\epsilon < 1$ 
(we restrict ourselves to the positive values of mass) the eigenvalues are real, while for 
$\epsilon > 1$ they are pure imaginary.

After one more transformation
\begin{eqnarray}
Z = P(x)W ,\nonumber\\
\nonumber\\
W^{\prime} = (P^{-1}CP - P^{-1}P^{\prime})W = DW ,\nonumber\\
\nonumber\\
P = I + \sum_{r=1}^{\infty}P_r x^{-r} ,\nonumber\\
\nonumber\\
D = C_0 + \sum_{r=1}^{\infty} D_r x^{-r} ,
\label{ZWPD}
\end{eqnarray}
where all matrices $D_r$ can be chosen to be diagonal, and the diagonal elements of matrices 
$P_r$ are all zero, the solution can written as follows
\begin{eqnarray}
Y &=& TP(x)\exp\left\{ \int\left( C_0 + \sum_{r=1}^{\infty} D_r x^{-r}\right)dx\right\} 
\nonumber \\
\nonumber \\
&=& 
T(I + F(x))\exp{C_0x + D_1\log x} ,
\label{asympmatr}
\end{eqnarray}
where the matrix $F(x)$ is holomorphic at $X = \infty$. All the matrices $P_r, D_r$ and, hence,
 $F$ can be calculated one by one from a recurrent equation readily derived from the above 
written expressions; they are of no importance for us. We are interested only in the matrix 
$D_1$ the equation for which reads as follows
\begin{equation}
D_1 = C_1 + C_0P_1 - P_1C_0 ,
\label{D1}
\end{equation}
and we are looking for the solution with $D_1$ having only diagonal elements nonzero and $P_1$ 
having zero diagonal elements. It is easy to see that the $D_1$ consists of diagonal elements 
if the matrix $C_1$, hence,
\begin{equation}
D_1 = (-1)^N(2N)!\frac{\alpha}{2} \mbox{diag}(\tilde t_{1,2N},\tilde t_{2,2N},...,
\tilde t_{2N,2N}) ,
\label{D1again}
\end{equation}
so the leading term in the first line of the matrix $Y$ (remember we need only these elements) 
is 
\begin{eqnarray}
\Psi_s = e^{\lambda_sx} e^{(-1)^N(2N)!(\alpha/2)\tilde t_{s,2N}\log x} ,\\
s = 1,2,...,2N\nonumber
\label{psis}
\end{eqnarray}
where $s$ is numerating linearly independent fundamental solutions.

The only thing left is to calculate the second exponent in Eqn.(\ref{psis}) and take a limit 
$N\to\infty$. Without loss of generality we can do this for only one value of $s$, let it be 
$s=1$. Making use of a specific structure of similarity matrix $T$, Eqn.(\ref{Ts}), we get for 
$\tilde t_{1,2N}$
\begin{equation}
\tilde t_{1,2N} = \frac{1}{(\lambda_1 - \lambda_2)(\lambda_1 - \lambda_3)\ldots(\lambda_1 - 
\lambda_{2N})} .
\label{tildet}
\end{equation}
But from Eqn.(\ref{lambda}) it follows that
\begin{eqnarray}
R_{2N} \equiv \sum_{K=0}^N \frac{(-1)^K \lambda^{2K}}{(2K)!} - \epsilon = 
\frac{(-1)^N}{(2N)!}(\lambda - \lambda_1)(\lambda - \lambda_2)\ldots(\lambda - \lambda_{2N}) ,
\nonumber\\
\nonumber\\
\lim_{\lambda\to\lambda_1} \frac{R_{2N}(\lambda)}{\lambda - \lambda_1} = 
R_{2N}^{\prime}(\lambda_1) = \frac{(-1)^N}{(2N)!}(\lambda_1 - \lambda_2)
(\lambda_1 - \lambda_3)\ldots(\lambda_1 - \lambda_{2N}) . 
\label{limit}
\end{eqnarray} 
Now it easy to take the limit $N\to\infty$. We get finally
\begin{eqnarray}
\Psi = e^{\lambda x} e^{-\frac{\alpha}{2\sin\lambda}\log x},\\
\nonumber\\
\cos{\lambda} = \epsilon .\nonumber
\label{psifinal}
\end{eqnarray}
For $\epsilon > 1$ the eigenvalues are pure imaginary, $\lambda = \pm i|\lambda|$, and we have 
ingoing and outgoing waves at infinity. For $\epsilon < 1$ the eigenvalues are real,
\begin{equation}
\lambda = \pm\cos^{-1}\epsilon + 2k\pi ,\qquad k = 0,\pm 1,\pm 2,...
\label{eigenv}
\end{equation}

Note that our original equation in finite differences has the following interesting property. 
Given the solution, say $\Psi_0$, we can construct new solution, multiplying $\Psi_0$ by 
any function $C(x)$ which is periodical with the period equal to the imaginary unit, $i$. 
That is,
\begin{eqnarray}
\Psi_0(x) \qquad\mbox{ - solution}\nonumber\\
\nonumber\\
\Psi_1(x) = C(x) \Psi_0(x) \qquad\mbox{ - solution, if}\nonumber\\
\nonumber\\
C(x+i) = C(x) .
\label{property}
\end{eqnarray}
\newpage
{\large \bf V. Solution of the Schroedinger Equation in Momentum Representation.}
\vskip1cm
It is convenient to introduce the following parameters, $\lambda$ and $\beta$,
\begin{equation}
\epsilon = \cos\lambda,\qquad \alpha = \kappa M^2 - e^2 = 2\beta\sin\lambda ,\nonumber
\end{equation}
in terms of which our Schroedinger equation becomes
\begin{equation}
\Psi(x+i) + \Psi(x-i) = 2\left(\cos\lambda + \frac{\beta\sin\lambda}{x}\right) \Psi(x) .
\label{main}
\end{equation}
Since for the black holes $\alpha > 0$, the signs of $\lambda$ and $beta$ are the same. 
We choose $\lambda > 0$ and, hence, $\beta > 0$.

Let us return to the operator form of the equation,
\begin{equation}
\left(\cosh\hat p - \beta \sin\lambda \hat x^{-1}\right) \Psi = \cos\lambda \Psi .
\label{operatoreqn}
\end{equation}
In the momentum representation operators $\hat p$ and $\hat x$ act as follows
\begin{equation}
\hat p \Psi_p = p \Psi_p ,\qquad \hat x \Psi_p = i \frac{\partial}{\partial p} \Psi_p ,
\label{operpx}
\end{equation}
where$|Psi_p$ is a wave function in the momentum representation.

The operator $\hat x^{-1}$ is not well defined. The ambiguity can be removed by adding 
suitable terms to the potential which are proportional to $\delta$-function and its 
derivatives at the origin. But, instead, we can multiply the equation by operator $\hat x$ 
from the left. By doing this we get
\begin{equation}
i\frac{\partial}{\partial p}(\cosh p - \cos\lambda)\Psi_p = \beta\sin\lambda \Psi_p ,
\label{dp}
\end{equation}
or
\begin{equation}
\frac{\partial}{\partial p} \log \Psi_p = - \frac{\sinh p + i\beta\sin\lambda}
{\cosh p - \cos\lambda} .
\label{logpsi}
\end{equation}
After introducing a new variable, $z = e^p$, the Eqn.(\ref{logpsi}) takes the form
\begin{eqnarray}
\frac{\partial}{\partial z} log \Psi_p = - \frac{z^2 + 2i\beta\sin\lambda z _ 1}
{z(z^2 - 2\cos\lambda z + 1)} = \frac{1}{z} - \frac{\beta + 1}{z - z_0} + 
\frac{\beta - 1}{z - \bar z_0} ,\nonumber\\
\nonumber\\
\mbox{where} \qquad z_0 = e^{i\lambda},\qquad \bar z_0 = e^{-i\lambda} .
\label{dzlog}
\end{eqnarray}

The above equation can be easily solved, the result is
\begin{eqnarray}
\Psi_p = C \frac{z}{(z - z_0)(z - \bar z_0)} 
\left(\frac{z - \bar z_0}{z - z_0}\right)^{\beta} ,\\
\nonumber\\
z = e^p .\nonumber
\label{solutionp}
\end{eqnarray}

This solution has very important property, it is periodical with the pure imaginary period 
$2\pi i$. This property will be explored in the next Section.
\newpage
{\large\bf VI. Transition to the Coordinate Representation. Fundamental Solution.}
\vskip1cm
In this Section we will transform the solution found in the momentum representation, to the 
coordinate representation. It can be done by the inverse Fourier transform,
\begin{equation}
\Psi(x) = \frac{1}{\sqrt{2\pi}} \int\limits_{-\infty}^{\infty} e^{ipx}\Psi_p dp .
\label{inverse}
\end{equation}

In the previous Section we obtained the unique (up to the multiplicative constant) solution 
for $\Psi_p$ , namely, Eqn.(\ref{solutionp}). But due to the periodicity of $\Psi_p$ we can 
shift the argument $p\to p+2\pi ki (k=0,\pm 1,\pm 2,...)$  which will result in the shift of 
the path of integration in the complex momentum plane from the real axis to the parallel one. 
And after such a shift we again obtain a solution in the coordinate representation. But,
\begin{eqnarray}
\Psi_1(x) = \frac{1}{\sqrt{2\pi}}\int\limits_{-\infty}^{\infty} e^{ipx}\Psi_p dp ,\nonumber\\
\nonumber\\
\Psi_2(x) = \frac{1}{\sqrt{2\pi}}\int\limits_{-\infty}^{\infty} e^{i(p+2\pi ki)x} 
\Psi_{p+2\pi ki} dp = \nonumber\\
e^{-2\pi kx} \frac{1}{\sqrt{2\pi}}\int\limits_{-\infty}^{\infty} e^{ipx}\Psi_p dp,\nonumber\\
\nonumber\\
\Psi_2(x) = e^{-2\pi kx} \Psi_1(x) .
\label{shifts}
\end{eqnarray} 

Thus, given one solution (say, $\Psi_0$), we can construct in this way a countable number of solutions
solutions, and, in general,
\begin{equation}
\Psi_{general}(x) = \left(\sum_{k=-\infty}^{\infty} c_k e^{-2\pi kx}\right) \Psi_0(x) .
\label{fourier}
\end{equation}
The infinite sum in parenthesis is nothing more but a Fourier series for a periodical 
function with an imaginary unit period $i$. In this way we reproduced the property of the 
solutions in coordinate representation stated in Sect.4.

How many different solutions do we need to construct the general solution? The very fact 
that in the momentum representation we obtained essentially one solution proves that for 
this we need only one solution $\Psi_0$ which we call the fundamental solution (though it 
deserves the name superfundamental). In the rest of this chapter we will construct the 
particular fundamental solution by the suitable choice of the contour of integration in the 
inverse Fourier transform integral.

Note, first of all, that our solution, Eqn.(\ref{inverse}), in momentum representation has 
countable number on branching points in the complex momentum plane which can be combined 
in pairs, $(i\lambda+2\pi ki, -i\lambda+2\pi(k+1)i), k=0,\pm 1,\pm 2,...$. Connecting two 
branching points in each pair by a cut we obtain the complex plane with countable number of 
cuts. On the corresponding Riemann surface our solution is a single valued analytical 
function of complex variable. 

Our choice of the contour of integration is as follows. We will integrate first along the 
real axes from left to right (i.e., from $-\infty$ to $+\infty$), then along a short curve 
at the right infinity $(p\to p+2\pi i, p\to{+\infty})$, then along the straight line 
$y=2\pi i$, parallel to the real axis, from left to right, and finally along a short curve 
at the left infinity back to the real axis. The integration along the short curves at 
infinities between the straight lines gives zero contributions for positive values of $x$ 
(for negative $x$ we can choose the straight line $y=-2\pi i$ instead of $y=2\pi i$). The 
integration along each straight line gives us a solution the linear combination of which is 
again a solution. Thus, the inverse Fourier integral along such a closed contour gives us 
a solution. This contour can be distorted to become a contour around the cut 
$(\lambda i, (2\pi-\lambda)i)$  for $x>0$ (or around the cut 
$(-\lambda i, -(2\pi -\lambda)i)$ for $x<0$).

Thus,
\begin{equation}
\Psi_0(x) = \frac{1}{\sqrt{2\pi}}\oint\limits_{C_+} e^{ipx} \Psi_p dp , \qquad x > 0 , 
\label{psipos}
\end{equation}  

\begin{equation}
\Psi_0(x) = \frac{1}{\sqrt{2\pi}}\oint\limits_{C_-} e^{ipx} \Psi_p dp , \qquad x < 0 , 
\label{psineg}
\end{equation}
where
\begin{eqnarray}
\Psi_p = C \frac{z}{(z - z_0)(z - \bar z_0)} \left(\frac{z - \bar z_0}{z - z_0}
\right)^{\beta}\\
\nonumber\\
z = e^p,\qquad z_0 = e^{i\lambda},\qquad \bar z_0 = e^{-i\lambda}\nonumber
\label{psimoment}
\end{eqnarray}

In what follows we restrict ourselves to the case of positive $x$ only.

The above integral representation for $\Psi_0$ can be simplified if we integrate 
Eqn.(\ref{psipos}) by parts. Using the fact that
\begin{equation}
\frac{zdp}{(z-z_0)(z-\bar z_0)} = - \frac{1}{2i\sin\lambda} 
d\left(\log{\frac{z-\bar z_0}{z-z_0}}\right) ,
\end{equation}

we get
\begin{equation}
\Psi_0(x) = x \oint\limits_{C_+} e^{ipx} \left(\frac{z-\bar z_0}{z-z_0}\right)^{\beta} dp ,
\qquad x> 0 ,
\label{intrepr}
\end{equation}
(the extra term vanishes because our contour has no boundary and for convenience we omitted 
some constant factors).

We will use use Eqn.(\ref{intrepr}) as an integral representation for the fundamental 
solution of our finite differences equation, Eqn.({main}).
\newpage
{\large\bf VII. Investigation of the Fundamental Solution.}
\vskip1cm
Let us investigate the fundamental solution, Eqn.(\ref{intrepr}), in 
more details. Our aim is to reduce the integration along the closed 
contour $C+$ around the cut to the finite interval between the 
corresponding branching points. But for $\beta\ge 1$ (remember that we 
have chosen $\beta$ positive) we have the nonintegrable singularity at 
the lower integration limit. To avoid this difficulty we need some 
recurrent relation for lowering the parameter $\beta$. To find such a 
relation we integrate by parts Eqn.(\ref{intrepr}) in the following way.
\begin{eqnarray}
\Psi_{\beta}(x) = x \oint\limits_{C_+} e^{ipx} \left(\frac
{z-\bar z_0}{z-z_0}\right)^{\beta} dp &=& \nonumber\\
x \oint\limits_{C_+} e^{ip(x+i)} (z-\bar z_0)^{\beta} 
\frac{z dp}{(z-z_0)^\beta} &=& \nonumber\\
\frac{x}{\beta -1} \oint\limits_{C_+} \left(e^{ip(x+i)}
(z-\bar z_0)^{\beta}\right)^{\prime} \frac{dp}{(z-z_0)^{\beta -1}} &=& 
\nonumber\\
\frac{x}{\beta -1}\left\{ \beta\oint\limits_{C_+} e^{ipx} 
\left(\frac{z-\bar z_0}{z-z_0}\right)^{\beta -1} dp \right.&+&  \nonumber \\
 i(x+i)
\oint\limits_{C_+}e^{ipx}\left(\frac{z-\bar z_0}{z-z_0}
\right)^{\beta -1}dp &-& \nonumber \\
\left.  i(x+i)\bar z_0 \oint\limits_{C_+}
e^{ip(x+i)} \left(\frac{z-\bar z_0}{z-z_0}\right)^{\beta -1} dp
\right\} &=& \nonumber\\
\frac{i}{\beta -1}\left\{ x - i(\beta -1) \Psi_{\beta -1}(x) - 
e^{-i\lambda x} \Psi_{\beta -1}(x+i)\right\}
\label{byparts}
\end{eqnarray}
Finally, we get
\begin{equation}
\Psi_{\beta}(x) = \Psi_{\beta -1}(x) + \frac{ix}{\beta -1}\left\{
\Psi_{beta -1}(x) - e^{-i\lambda} \Psi_{\beta -1}(x+i)\right\} .
\label{recurrent}
\end{equation}

From the above relation it is easy to derive the structure of 
$\Psi_{\beta}$ for general values of $\beta >1$. Let us take 
$\beta =n+\tilde\beta$, with $\tilde\beta \le 1$. Then, $\Psi_{\beta}$ is 
the sum of two terms, which of them is the product of some polynomial of 
$n$-th degree and of $\Psi_{\tilde\beta(x)}$ or $\Psi_{tilde\beta}(x+i)$. 
Thus, we can proceed assuming $\beta\le 1$.

First of all, consider separately the case $\beta =1$. We have
\begin{eqnarray}
\Psi_1 = x \oint\limits_{C_+} e^{ipx}\frac{e^p - e^{-i\lambda}}
{e^p - e^{i\lambda}} dp &=& 
2\pi ix \lim_{p\to i\lambda} \frac{e^{ipx}(e^p - e^{-i\lambda})
(p - i\lambda)}{e^p - e^{i\lambda}} \nonumber\\
&=& - \left(4\pi e^{-i\lambda}\sin\lambda\right) xe^{-\lambda x} .
\label{beta1}
\end{eqnarray}
From this it follows that all $\Psi_{\beta}$ for positive integer 
$\beta =n$ has the form
\begin{equation}
\Psi_n = P_n(x) e^{-\lambda x} ,
\label{psin}
\end{equation} 
where $P_n(x)$ are some polynomials of $n$-th degree. These polynomials 
were first introduced in \cite{me} .

In the case $0<\beta <1$ the integrand in the Eqn.(\ref{intrepr}) is 
integrable on both ends of the cut in the complex momentum plane and we 
are able to convert the integral along the closed contour $C_+$ into the 
integral along the finite interval. We get the following,
\begin{equation}
\Psi_{\beta}(x) = \oint\limits_{C_+} e^{ipx} f_p dp = \left(1 - 
e^{2\pi i\beta}\right) x \Phi_{\beta}(x) ,
\end{equation}
where
\begin{equation}
\Phi_{\beta}(x) = \int\limits_{i\lambda}^{i(2\pi -\lambda}) e^{ipx}
\left(\frac{e^p - e^{-i\lambda}}{e^p - e^{i\lambda}}\right)^{\beta} dp ,
\qquad x > 0 .
\label{phi}
\end{equation}

Changing the variables,
\begin{equation}
e^q - e^{i\lambda} = -(2i\sin{\lambda}) y ,
\label{qy}
\end{equation}
we get
\begin{equation}
\Phi_{\beta} = \left(1 - e^{-2i\lambda} \right) e^{-2\lambda x} \int\limits_0^1 
\left(1 - \left(1-e^{-2i\lambda}\right) y \right)^{ix-1} \left(
\frac{y - 1}{y}\right)^{\beta} dy .
\label{phibeta}
\end{equation}
Comparing this with the well known integral representation for the Gauss's hypergeometric 
function, $F(a,b;c;z)$, \cite{Ryzhik} ,
\begin{eqnarray}
F(a,b;c;z) = \frac{1}{B(b,c-b)} \int\limits_0^1 t^{b-1} (1 - t)^{c-b-1} (1 - tz)^{-a} dt ,
\nonumber\\
B(x,y) = \int\limits_0^1 t^{x-1} (1 - t)^{y-1} dt ,
\label{Gauss}
\end{eqnarray}
we see that the integral in Eqn.(\ref{phibeta}) is the Gauss's hypergeometric function 
$F(a,b;c;z)$ with the following values of parameters,
\begin{equation}
a = 1 - ix,\qquad b = 1 - \beta;\qquad c = 2;\qquad z = 1 - e^{-2i\lambda} , 
\label{param}
\end{equation}
and
\begin{equation}
B(b,c-b) = B(1-\beta,1+\beta) = \frac{\pi \beta}{\sin{\pi \beta}} .
\label{B}
\end{equation}

Finally, for the fundamental solution $\Psi_{\beta}(x)$ we get
\begin{equation}
\Psi_{\beta}(x) = \left(-4\pi\beta e^{-i\lambda} \sin\lambda\right) x e^{-\lambda x} 
F\left(1 - ix, 1 - \beta; 2; 1 - e^{-2i\lambda}\right) .
\label{fundhyper}
\end{equation}

The above expression was derived for $0 < \beta , 1$ only. We shaw now that the same is 
valid for any value $\beta > 0$.

Let us check first the relation for $\beta = 1$. Indeed, from Eqn(\ref{fundhyper}) it 
follows that
\begin{equation}
\Psi_1 = -\left(4\pi e^{-i\lambda}\sin\lambda\right) x e^{-\lambda x} F(1-ix,0;2;
1-e^{-2i\lambda}) .
\label{psione}
\end{equation}
But $F(a,0;c;z)=1$ and we see that Eqn.(\ref{psione}) coincides with Eqn(\ref{beta1}) 
derived directly from our integral representation.

Second, we show now that the recurrent relations, Eqn.(\ref{recurrent}), for lowering index 
$\beta$ are direct consequences of the properties of Gauss's hypergeometric function.
Indeed, let us write the following two of the well known Gauss's recurrent relations between 
hypergeometric functions with different values of parameters \cite{Ryzhik} ,
\begin{eqnarray}
c(c-bz-a) F(a,b;c;z) - c(c-a) F(a-1,b;c;z) 
\nonumber \\
\nonumber \\
+ abz(1-z) F(a+1,b+1;c+1;z) = 0
\nonumber \\
\nonumber\\
c(c-az-b) F(a,b;c;z) - c(c-b) F(a,b-1;c;z) 
\nonumber \\
\nonumber \\
+ abz(1-z) F(a+1,b+1;c+1;z) = 0\nonumber .
\end{eqnarray}
Subtracting one from another, we get
\begin{equation}
(c-b) F(a,b-1;c;z) = (a-b)(1-z) F(a,b;c;z) + (c-a) F(a-1,b;c;z) .
\end{equation}
Substituting our values of parameters we obtain
\begin{eqnarray}
\beta F(1-ix,1-\beta;2;1-e^{-2i\lambda} =\\
(\beta -ix-1) e^{-2i\lambda} F(1-ix,1-(\beta -1);2;1-e^{-2i\lambda}) 
\nonumber \\
+ (1+ix) F(1-i(x-i),
1-(\beta -1);2;1-e^{-2i\lambda}) .\nonumber
\end{eqnarray}
And, in terms of $\Psi_{\beta}$ we have
\begin{equation}
\Psi_{\beta}(x) = e^{-2i\lambda} \left\{\Psi_{\beta -1}(x) - \frac{ix}{\beta -1}
\left(\Psi_{\beta -1}(x) - e^{i\lambda} \Psi_{\beta -1}(x-i)\right)\right\}\nonumber .
\end{equation}
Using the original finite differences equation we can rewrite this in the form
\begin{equation}
\Psi_{\beta}(x) = \Psi_{\beta -1}(x) + \frac{ix}{\beta -1} \left(\Psi_{\beta -1}(x)
 - e^{-i\lambda} \Psi_{\beta -1}(x+i)\right)
\label{recurrentnew}
\end{equation}
which is just the recurrent relation, Eqn.(\ref{recurrent}), derived directly from the 
integral representation.

And, finally, it can be easily shown that the original equation in finite differences. 
Eqn.(\ref{main}), is a direct consequence of the following Gauss's recurrent relation 
\cite{Ryzhik} ,
\begin{equation}
(2a-c-az+bz) F(a,b;c;z) + (c-a) F(a-1,b;c;z) + a(z-1) F(a+1,b;c;z) = 0\nonumber
\end{equation}

At the end of this section we consider the asymptotics of the general solution at 
$x \to 0$ and $x \to \infty$ .

From the hypergeometric expansion \cite{Ryzhik}
\begin{equation}
F(a,b;c;z) = 1 + \sum_{k=1}^{\infty} \frac{a(a+1)\ldots(a+k-1)b(b+1)\ldots(b+k-1)}
{c(c+1)\ldots(c+k-1) k!} z^k\nonumber 
\end{equation}
and the form of the general solution
\begin{eqnarray}
\Psi_{gen} = \Psi_0(x) \sum_{k=-\infty}^{\infty} c_k e^{-2\pi x}\nonumber\\
\nonumber\\
\Psi_0 \propto x e^{-\lambda x} F(1-ix, 1-\beta; 2; 1-e{-2i\lambda})\nonumber
\end{eqnarray}
it follows immediately that the asymptotics at $x\to 0$ are
\begin{equation}
x^r,\qquad r = 1,2... .
\end{equation}
This is just what we obtained earlier in Sect.IV.

To obtain the asymptotics at $x\to \infty$ we make use of the following transformation 
relation for the hypergeometric function \cite{Ryzhik}
\begin{eqnarray}
F(a,b;c;z) = \frac{\Gamma(c)\Gamma(c-a-b)}{\Gamma(c-a)\Gamma(c-b)}F(a,b;a+b-c+1;1-z) +
\nonumber\\
\nonumber\\ 
(1-z)^{c-a-b}\frac{\Gamma(c)\Gamma(a+b-c)}{\Gamma(a)\Gamma(b)}F(c-a,c-b;c-a-b+1;1-z)
\nonumber ,
\end{eqnarray}
where $\Gamma(\ldots)$ is the Euler's function. We get,then,
\begin{eqnarray}
\Psi_{\beta} &=& \left(-4\pi\beta e^{-i\lambda(\beta +1)}\sin\lambda\right) x \nonumber \\
&&\left\{\frac{\Gamma(ix+\beta)e^{i\lambda\beta} e^{-\lambda x}}{\Gamma(1+ix)\Gamma(1+\beta)} 
 F(1-ix,1-\beta;1-ix-\beta;e^{-2i\lambda})\right. 
\nonumber \\
&+& 
\left. \frac{\Gamma(-ix-\beta)e^{-i\lambda\beta} e^{\lambda x}}{\Gamma(1-ix)\Gamma(1-\beta)} 
 F(1+ix,1+\beta;1+ix+\beta;e^{-2i\lambda})\right\}
\label{infty}
\end{eqnarray}
Writing the hypergeometric function in the form
\begin{equation}
F(a,b;c;z) = \frac{\Gamma(c)}{\Gamma(a)\Gamma(b)} \sum_{k=0}^{\infty} 
\frac{\Gamma(a+k)\Gamma(b+k)}{\Gamma(c+k) k!} z^k\nonumber
\end{equation}   
and using the relation
\begin{equation}
\Gamma(z)\Gamma(1-z) = \frac{\pi}{\sin{\pi z}}\nonumber
\end{equation}

we can rewrite $\Psi_{\beta}$ in the following way
\begin{eqnarray}
\Psi_{\beta} &=& \frac{4\pi\beta e^{-i\lambda(\beta+1)}\sin{\lambda}}
{\Gamma(1-\beta)\Gamma(1+\beta)} \frac{e^{\pi x} - e^{-\pi x}}{e^{\pi x-i\pi\beta} 
- e^{-\pi x+i\pi\beta}} 
\nonumber \\ 
&&\left\{ e^{i\lambda\beta} e^{-\lambda x} \sum_{k=0}^{\infty}
\frac{\Gamma(1-ix+k)\Gamma(1+k-\beta)}{\Gamma(1+k-ix-\beta) k!} e^{-2i\lambda k} \right.
\nonumber \\
&-&
\left. e^{-i\lambda\beta} e^{\lambda x} \sum_{k=0}^{\infty}
\frac{\Gamma(1+ix+k)\Gamma(1+k+\beta)}{\Gamma(1+k+ix+\beta) k!} e^{-2i\lambda k}\right\}
\nonumber
\end{eqnarray}

Using then the asymptotic form of the Euler function,
\begin{equation}
\Gamma(z) = z^{z-\frac{1}{2}} \sqrt{2\pi} \phi(\frac{1}{z}),\qquad z \to \infty\nonumber
\end{equation}

we find the asymptotic behavior of our fundamental solution $\Psi_{\beta}(x)$,
\begin{eqnarray}
\Psi_{\beta} = \frac{4\pi\beta e^{-i\lambda(\beta+1)} e^{i\pi\beta} \sin{\lambda}}
{\Gamma(1-\beta)\Gamma(1+\beta)} \left\{ e^{i\lambda\beta} e^{-\lambda x} 
e^{\frac{i\pi\beta}{2}} \sum_{k=0}^{\infty}
\frac{\Gamma(1+k-\beta)}{k!} e^{-2i\lambda k} \right. \nonumber \\ -
\left. e^{-i\lambda\beta} e^{\lambda x} e^{\frac{-i\pi\beta}{2}}
\sum_{k=0}^{\infty} \frac{\Gamma(1+k+\beta)}{k!} e^{-2i\lambda k}\right\}\nonumber
\end{eqnarray}

But
\begin{eqnarray}
\sum_{k=0}^{\infty} \frac{\Gamma(1+k-\beta)}{k!} e^{-2i\lambda k} = \Gamma(1-\beta)
(1-e^{-2i\lambda})^{\beta -1} 
\nonumber \\ 
= \Gamma(1-\beta)e^{-i\lambda(\beta-1)}2^{\beta-1}
e^{\frac{i\pi}{2}(\beta-1)}\sin{\lambda}^{\beta-1} ,\nonumber
\end{eqnarray}

and we obtain finally
\begin{eqnarray}
\Psi_{\beta} &=& - 2\pi i \beta e^{-i\lambda\beta} e^{i\pi\beta} \left\{
\frac{(2\sin{\lambda})^{\beta}}{\Gamma(1+\beta)} x^{\beta} e^{-\lambda x} -
\frac{(2\sin{\lambda})^{-\beta}}{\Gamma(1-\beta)} x^{-\beta} e^{\lambda x}\right\}
\phi(\frac{1}{x}) ~,
\nonumber \\
 x &\to& \infty 
\label{inftyasymp}
\end{eqnarray}
From this it follows that the asymptotic behavior of the general solution is exactly 
the same as was found earlier.
\newpage
{\large\bf VIII. Boundary Conditions and Conserved Current.}
\vskip1cm
In the preceding sections we found the general solution to the equation in finite 
differences with Coulomb potential,
\begin{equation}
\Psi(x + i) + \Psi(x - i ) = 2 \left(\cos{\lambda} + \frac{\beta}{x} \sin{\lambda}\right) 
\Psi(x) .
\label{ours}
\end{equation} 
Now we should remember that this equation is actually the radial 
Schroedin\-ger equation for 
zero angular momentum, and the coordinate $x$ runs from zero to infinity, i.e., $x$ takes 
values in the positive semi-axis only. Thus the Hamiltonian,
\begin{equation}
\hat H = \cosh{\left( i\frac{\partial}{\partial x}\right)} - \frac{\beta}{2x} \sin{\lambda}
\label{Hamilt}
\end{equation}
should be the selfadjoint operator on the positive semi-axis rather than on the whole real 
axis as dictated by the quantum mechanics postulates. In addition, the wave function should 
be square integrable on the semi-axis.
The corresponding extension of the above Hamiltonian was found by P.Hajicek in 
\cite{Hajicek}.

It appeared the wave function should obey the following boundary conditions at the origin, 
$x = 0$ ,
\begin{equation}
\Psi^{(2n)}(0) = 0, \qquad  n = 0,1,...  .
\label{boundorigin}
\end{equation}
That is, the function itself and all its even derivatives should be zero at the origin.
 
The appearance of infinite number of conditions is due to the infinite order of the 
operator.

P.Hajicek found also a conserved probability current $J(x)$ for the 
Schroe\-dinger equation in 
finite differences, Eqn.(\ref{ours}) . It is given as usual by the equation
\begin{equation}
J^{\prime}(x) = i \left(\Psi^* {\hat H} \Psi - \Psi {\hat H} \Psi^* \right)\nonumber
\end{equation} 
so that 
\begin{equation}
\left( \Psi^{*} \Psi \right)^. + J^{\prime} = 0\nonumber
\end{equation}
Writing the Hamiltonian (\ref{Hamilt}) as follows
\begin{equation}
\hat H = \frac{1}{2} \sum_{k=0}^{\infty} \frac{(-1)^k}{(2k)!} \Psi^{(2k)}(x) - 
\frac{\beta\sin{\lambda}}{2x}
\label{Hamiltser}
\end{equation}   
we obtain for $J(x)$ ,
\begin{equation}
J(x) = i \sum_{k=0}^{\infty} \frac{(-1)^k}{(2k)!} \sum_{l=1}^k (-1)^{l-1}
\left[\Psi^{*(l-1)} \Psi{(2k-l)} - \Psi^{(l-1)} \Psi^{*(2k-l)}\right] .
\label{currenthaj}
\end{equation} 

P.Hajicek showed that the boundary condition (\ref{boundorigin}) implies that
\begin{equation}
J(0) = 0 .
\label{currentzero}
\end{equation}

The boundary condition (\ref{boundorigin}) is a direct generalization of the boundary 
condition in the case of nonrelativistic Schroedinger equation which is of second order. In 
this particular case the boundary conditions (\ref{boundorigin}) are reduced to the single 
one, $\Psi(0)=0$ , which is together with the square integrability condition ensures 
existence of a discrete spectrum for bound states.
And, in general, let us remind the results of Sect.IV, where we considered the truncated 
Hamiltonian of $2N$th order. It was shown that the asymptotics at the origin are
\begin{equation}
\Psi \sim x , x^2 , ... , x^{2N-1} , 1 + x^{2N-1} \log x 
\label{asymptrunc}
\end{equation}
and at the infinity they are of the form
\begin{equation}
\Psi \sim e^{\pm \lambda_k x} e^{\alpha_k\log x} ,\qquad k=1,...N .
\label{asyptrunc2}
\end{equation}
where $\lambda_k$ are real for bound states. The general solution is defined up to a 
normalization factor and depends, actually, on $(2N-1)$ arbitrary constants which are to 
be determined using boundary conditions. The square integrability condition reduces the 
number of unknown constants to $(N-1)$. But for the truncated operator of $2N$th order 
we have $N$ conditions at the origin. They can only be satisfied for a discrete spectrum 
of eigenvalues.

Note that the last of the asymptotics (\ref{asymptrunc}) is excluded by the boundary 
condition at the origin.

We can reverse the above consideration and start from the asymptotic behavior at the 
origin. To satisfy the boundary conditions there we have to exclude the last asymptotics 
(\ref{asymptrunc}) and all the even terms in the expansion up to the $(2N-2)$th order. 
By doing this can determine $N$ of $(2N-1)$ unknown constants. Thus we are left with only 
$(N-1)$ constants for N boundary conditions at infinity.
But the situation is not so simple in the case of an infinite order operator.
First of all, how to get rid of the logarithmic term in asymptotics? The problem is that if 
we first go to the infinite order limit, then the logarithmic term unconditionally 
disappears $(|x|<1 !)$ leaving $1$ as an asymptotics. But the latter can be easily 
reproduced by some infinite linear superposition of the remained asymptotics. Subtracting 
one from another we would obtain the solution which would satisfy formally all the boundary 
conditions at the origin still having enough arbitrary constants to satisfy boundary 
conditions at infinity.

But if we first differentiate $2N$ times the asymptotic solution 
containing logarithmic term 
we will get ($1/x$) term. Thus, requiring that the ``good'' solution should be infinitely 
differentiable we can reach our goal. Moreover, we have to require the analyticity of the 
wave function at least on the real axis. Thus is because we implicitly have used the 
analyticity of the solutions in transition from the differential equation of infinite order 
to the finite differences equation.
The situation is not good at the infinity either. We saw that in the case of our particular 
equation we can obtain a new solution multiplying the known one by $\exp{(-2\pi kx)}$ . 
Choosing the large enough value of $k$ we are able to convert the ``bad'' at infinity 
solution to the ``good'' one.

We will see in the next section that it is the analyticity condition which solves this 
problem.
\newpage
{\large \bf IX. Discrete Spectrum. Polynomials and Generating Function.}
\vskip1cm
We have shown in Sect.VII that the fundamental solution 
\begin{equation}
\Psi_{\beta}(x) = \left(-4\pi\beta e^{-i\lambda} \sin\lambda\right) x e^{-\lambda x} 
F\left(1 - ix, 1 - \beta; 2; 1 - e^{-2i\lambda}\right) .
\label{fundsol}
\end{equation}
can be written in the form
\begin{eqnarray} 
\Psi_{\beta} &=& \left(-4\pi\beta e^{-i\lambda(\beta +1)}\sin\lambda\right) x 
\nonumber \\
&&\left\{ \frac{\Gamma(ix+\beta)e^{i\lambda\beta} e^{-\lambda x}}
{\Gamma(1+ix)\Gamma(1+\beta)} 
 F(1-ix,1-\beta;1-ix-\beta;e^{-2i\lambda} \right. \nonumber \\
&+& \left.
\frac{\Gamma(-ix-\beta)e^{-i\lambda\beta} e^{\lambda x}}{\Gamma(1-ix)\Gamma(1-\beta)} 
 F(1+ix,1+\beta;1+ix+\beta;e^{-2i\lambda})\right\}
\label{soltwogamma}
\end{eqnarray}
and the asymptotic behavior at $x\to\infty$ is the following 
\begin{eqnarray}
\Psi_{beta} &=& - 2\pi i \beta e^{-i\lambda\beta} e^{i\pi\beta} \left\{
\frac{(2\sin{\lambda})^{\beta}}{\Gamma(1+\beta} x^{\beta} e^{-\lambda x} -
\frac{(2\sin{\lambda})^{-\beta}}{\Gamma(1-\beta} x^{-\beta} e^{\lambda x}\right\}
\phi(\frac{1}{x}) ~,\nonumber \\
x &\to& \infty 
\label{inftyagain}
\end{eqnarray}
From this it is clear that the fundamental solution is an analytic function only if
\begin{equation}
\beta = n
\label{betan}
\end{equation}
(remember that we chose $\beta >0$).
This leads to the discrete spectrum for the total mass $m$. Indeed,we have
\begin{eqnarray}
\beta = \frac{\alpha}{2\sin{\lambda}} = n ,\qquad \alpha = \kappa M^2 ,\nonumber\\
\epsilon = \frac{m}{M} = \cos{\lambda} ,\nonumber\\
\epsilon = \sqrt{1 - \frac{\alpha^2}{4n^2}}
\label{epsilonn}
\end{eqnarray}
For $\beta=n$ the hypergeometric series in Eqn(\ref{fundsol}) terminates and we are left 
with the following fundamental solution,
\begin{equation}
\Psi_n(x) = P_n(x) e^{-\lambda x} ,
\label{solpoly}
\end{equation}
where $P_n$'s are some polynomials of order $n$. They can be calculated directly from the 
definition,
\begin{equation}
P_n(x) = \left(-4\pi n e^{-i\lambda}\sin{\lambda}\right) x e^{-\lambda x} 
F(1-ix,1-n;2;1-e^{-2i\lambda}) .
\label{defpoly}
\end{equation}
   We can also make use of the recurrent relation found earlier which for our polynomials 
$P_n$ takes the form
\begin{equation}
P_{n+1}(x) = P_n(x) + \frac{ix}{n} \left\{P_n(x) - e^{-2i\lambda} P_n(x+i)\right\} .
\label{recurpoly}
\end{equation}
In this way we are able to calculate $P_n(x)$ step by step, starting from $P_1$ :
\begin{equation}
P_1(x) = \left(-4\pi e^{-i\lambda} \sin{\lambda}\right) x .
\label{poly1}
\end{equation}
But there exists yet another way to derive polynomials (\ref{solpoly}) . To find it we start 
with the following differential relation for hypergeometric function $F(a,b;c;z)$ 
\cite{Olver}
\begin{equation}
\left(\frac{\partial}{\partial z}\right)^k \left\{z^{b+k-1} F(a,b;c;z)\right\} = 
\frac{\Gamma(b+k)}{\Gamma(b)} z^{b-1} F(a,b+k;c;z) ,\nonumber
\end{equation}
which in terms of polynomials $P_n$ reads as follows,
\begin{equation}
\left( \frac{\partial}{\partial z}\right)^k \left\{ \frac{z^{k-n-1}}{k} P_n(x)\right\} = 
\frac{\Gamma(1+k-n)}{\Gamma(1-n)} \frac{z^{-n-1}}{n-k} P_{n-k}(x) ,\nonumber
\end{equation}
For $k=1$ it becomes (remember that $z=\exp{-2i\lambda}$)
\begin{equation}
\frac{\partial}{\partial\lambda} \left\{\frac{e^{i\lambda n}}{n(\sin{\lambda})^n} 
P_n(x)\right\} = - \frac{e^{i\lambda(n-1)}}{(\sin{\lambda})^{n+1}} P_{n-1}(x) .\nonumber
\end{equation}
Using the relation
\begin{equation}
P_{n-1}(x) = - e^{2i\lambda} P_{n+1}(x) + 2 e^{i\lambda} \left(\cos{\lambda} - 
\frac{x}{n}\sin{\lambda}\right) P_n(x)\nonumber
\end{equation}
which follows from the following Gauss's recurrent relation \cite{Ryzhik}
\begin{equation}
(c-b)F(a,b-1;c;z) = (2b-c+(a-b)z)F(a,b;c;z) + b(z-1)F(a,b+1;c;z) ,\nonumber
\end{equation}
we get
\begin{eqnarray}
\frac{\partial}{\partial\lambda} \left(\frac{e^{i\lambda n} P_n(x)}{n(\sin{\lambda})^n}
\right) &=& \frac{e^{i\lambda(n+1)}}{(\sin{\lambda})^{n+1}} P_{n-1}(x) 
\nonumber \\ &-& 
2\cot{\lambda}\frac{e^{i\lambda n}}{(\sin{\lambda})^n} P_n(x) + \frac{2x}{n}
\frac{e^{i\lambda n}}{(\sin{\lambda})^n} P_n(x) .\nonumber
\end{eqnarray}
After rescaling the polynomials
\begin{equation}
\tilde P_n(x) = (-1)^n\left(e^{i\lambda}\sin{\lambda}\right)^n P_n(x)\nonumber
\end{equation}
we arrive at the following relation
\begin{equation}
ne^{-3\lambda x} \tilde P_{n+1}(x) = \frac{\partial}{\partial\cot{\lambda}}
\left(e^{-2\lambda x} \tilde P_n(x)\right) \nonumber
\end{equation}
which finally results in
\begin{eqnarray}
e^{-2\lambda x} \tilde P_{n+1}(x) = \frac{1}{n!}
\left(\frac{\partial}{\partial\cot{\lambda}}\right)^n\left(e^{-2\lambda x} 
\tilde P_1(x)\right)\\
\nonumber\\
\tilde P_1(x) = (4\pi\sin^2{\lambda}) x .\nonumber
\label{pnp1}
\end{eqnarray}
Writing
\begin{equation}
e^{-2\lambda x} = \left(\frac{\cot{\lambda} + i}{\cot{\lambda} - i}\right)^{ix}\nonumber
\end{equation}
and noting that
\begin{equation}
(\sin{\lambda}^2) x e^{-2\lambda x} = \frac{1}{2}\frac{\partial}{\partial\cot{\lambda}}
\left(\frac{\cot{\lambda} + i}{\cot{\lambda} - i}\right)^{ix}\nonumber
\end{equation}
we get
\begin{equation}
\tilde P_n(x) = \frac{e^{-2\lambda x}n}{2}\frac{1}{n!}
\left(\frac{\partial}{\partial\cot{\lambda)}}\right)^n
\left(\frac{\cot{\lambda} + i}{\cot{\lambda} - i}\right)^{ix}\nonumber
\label{almostgen}
\end{equation}
From this it follows the form of the generating function for our (once more rescaled) 
polynomials $\Pi_n(x) = (1/n)\tilde P_n(x)$
\begin{eqnarray}
\Phi(z, x) = \sum_{n=0}^{\infty} \frac{\Phi^{(n)}(0, x)}{n!} z^n = 
\sum_{n=0}^{\infty} \Pi_n(x) z^n = \nonumber\\
2\pi e^{2\lambda x} \left( \frac{z + \cot{\lambda} + i}{z +\cot{\lambda} - i}\right)^{ix}
\label{genfunc}
\end{eqnarray}
\newpage
{\large\bf X. Wave Function of the Ground State.}
\vskip1cm
In this section we give an example of solving the boundary conditions and find the wave 
function of the ground state (i.e., for $n=1$).
In the preceding section the fundamental solution for $n=1$ was found to be 
\begin{equation}
\Psi_1 = x e^{-\lambda x}
\label{ps1}
\end{equation}
(we omitted the irrelevant here constant factor). The general solution obeying the 
boundary condition at infinity (exponential falloff at $x\to\infty$) has the form
\begin{equation}
\Psi_{1 gen} = x e^{-\lambda x} \sum_{k=0}^{\infty} c_k e^{-2\pi kx} ,
\label{ps1gen}
\end{equation}
and the coefficients $c_k$ are to be determined  by solving boundary conditions at the 
origin,
\begin{equation}
\Psi^{(2l)}(0) = 0,\qquad l = 0, 1, ... 
\label{orig}
\end{equation}
Differentiating Eqn.(\ref{ps1gen}) $2l$ times we get the following infinite set of linear 
equations, 
\begin{equation}
2l\sum_{k=0}^{\infty} c_k (\lambda + 2\pi k)^{2l-1} = 0, \qquad l=0,1,...
\label{linset}
\end{equation}
The determinant of this system is identically zero because the first line consists of 
zeros.

We can truncate this system at some specific value of $l$, calculate all the determinants 
and minors and then, after appropriate renormalization (say, putting $c_0=1$), take the 
limit $l\to\infty$. In our simplest case of the ground state it is possible to make all 
the calculations up to the very end with the following result
\begin{equation}
c_k = \frac{(-1)^k}{k!} \frac{\Gamma(\frac{\lambda}{\pi} + k)}
{\Gamma(\frac{\lambda}{\pi})} . \qquad k=0,1,...
\label{ck}
\end{equation}  
Inserting this into Eqn.(\ref{ps1gen}) we can write the ground state wave function 
(up to the normalization factor) in a very simple and nice form,
\begin{eqnarray}
\Psi_1 = x e^{-\lambda x} \sum_{k=0}^{\infty} c_k e^{-2\pi kx} =\nonumber\\
x e^{-\lambda x} \sum_{k=0}^{\infty} \frac{(-1)^k}{k!}\frac{\lambda}{\pi}
\left(\frac{\lambda}{\pi}+1\right)\ldots 
\left(\frac{\lambda}{\pi}+k-1\right)e^{-2\pi kx} =\nonumber\\
x e^{-\lambda x}\left(1 + e^{-2\pi x}\right)^{-\frac{\lambda}{\pi}} =\nonumber\\
2^{-\frac{\lambda}{\pi}}\frac{x}{(\cosh{\pi x})^{\frac{\lambda}{\pi}}} .
\label{wavegr}
\end{eqnarray}
It is easy to see that after shifting the parameter $\lambda\to\lambda+2\pi l$, where $l$ 
is a positive integer, we again obtain a solution satisfying all the boundary conditions 
(technically such a solution can be obtained if put zero first $l$ coefficients 
$c_m, m=0,...(l-1)$ in the infinite set of linear equations considered above and normalize 
to unity $c_l$). Moreover, the introduced earlier conserved current is identically zero 
for any superposition (with complex coefficients) of these functions.
This means that each eigenvalue of a total mass (energy) is infinitely degenerate. Clearly, 
the origin of such a degeneracy lies in freezing all but one dynamical degrees of freedom. 
Our experience in quantum mechanics suggests that after restoring the frozen degrees of 
freedom this degeneracy disappears. As a result we expect that the ``true'' spectrum for
thin shell masses will depend on more than one quantum number. Due to infinite degeneracy 
of the ``frozen'' spectrum the ``true'' one will, in fact, be quasicontinuous. And, as a 
consequence, the Hawking's evaporation spectrum will be quasicontinuous as well. The 
existence of the ground state with nonzero mass means that the evaporation process should 
stop when the mass of a black hole approaches the minimum possible value 
$(\approx 0.9 M_{Pl})$ . And we may conjecture that the final state of the black hole 
evaporation will be characterized by the minimal value of mass but different wave 
functions of a ground state depending on details of mass distribution and evaporation 
process. This unexpected feature of the black hole mass spectrum (its infinite degeneracy) 
reminds the problem of so called hidden parameters in quantum mechanics and may be helpful 
in resolving the well known information paradox in a black hole physics.
And at the end of this section we give the result of calculation for the norm of the 
ground state wave function. The scalar product of any two ground state wave function with 
parameters $\lambda_1$ and $\lambda_2$ is given by
\begin{eqnarray}
N_{\lambda_1,\lambda_2}^2 = 2^{-\frac{\lambda_1 +\lambda_2}{\pi}} \int\limits_0^{\infty} 
\frac{x^2 dx}{(\cosh{\pi x})^{\frac{\lambda_1 +\lambda_2}{\pi}}} =\nonumber\\
\frac{1}{4\pi^{5/2}} 2^{-\frac{\lambda_1 +\lambda_2}{\pi}} 
\frac{\Gamma\left(\frac{\lambda_1 +\lambda_2}{2\pi}\right)}
{\Gamma\left(\frac{1}{2} + \frac{\lambda_1 +\lambda_2}{2\pi}\right)} 
\zeta\left(2, \frac{\lambda_1 +\lambda_2}{2\pi}\right) ,
\label{product}
\end{eqnarray}
where $\zeta(x,y)$ is a Riemann's function of two variables. Thus, for a wave function 
with given value of $\lambda$ we get
\begin{equation}
N_{\lambda}^2 = \frac{1}{4\pi^{5/2}} 2^{-\frac{2\lambda}{\pi}} 
\frac{\Gamma\left(\frac{\lambda}{\pi}\right)}
{\Gamma\left(\frac{1}{2} + \frac{\lambda}{\pi}\right)} \zeta\left(2, \frac{\lambda}{\pi}
\right) .
\label{norm}
\end{equation}
\newpage
{\large\bf XI. Quantum Black Holes.}
\vskip1cm
In the preceding sections we have got a complete solution to the Schroe\-dinger equation in 
finite differences describing a quantum mechanical behavior of the self-gravitating 
electrically charged spherically symmetric dust shell. For the bound states we found the 
discrete spectrum of mass which coincides with Dirac spectrum if we put zero the so called 
the so called radial quantum number. Similarly to the case of hydrogen atom, to which our 
spectrum reduces in the nonrelativistic limit, we may conclude that the shell does not 
collapse without radiation. The starting classical situation is however different. The 
classical hydrogen atom collapses radiating continuously while the classical spherically 
symmetric shell collapses without radiation. But without such a collapse the event horizon 
and thus the black hole can not be formed at all. Moreover, if we calculate the mean value of
radius of the shell using the wave function we will find that at least for large values of 
principal quantum number it is far away of classical event horizon, so the shell spends most 
of its ``lifetime'' outside the classical black hole. So, to get a black hole solution we 
need some new criterium and more detailed investigation of the spectrum is needed.
Let us consider the dependence of a total mass $m$ on a bare mass $M$ , the other variables, 
$e$ and $n$ fixed. We see that there are two branches, an increasing one for
\begin{equation}
(\kappa M^2 - e^2)(3\kappa M^2 - e^2)<4n^2
\label{branch}
\end{equation}
and a decreasing one in the opposite case. Note that now
\begin{equation}
\frac{m}{M} > \sqrt{\frac{2}{3}} \sqrt{1+ \frac{e^2}{3\kappa M^2 - e^2}} ,
\label{nmineq}
\end{equation}
and this value is greater than the corresponding classical value.

The increasing branch corresponds to the ``black hole case'' while the decreasing branch - 
to the ``wormhole case''. Using a ``quasiclassical'' argumentation we can say that for 
the states obeying inequality (\ref{branch}) the expectation value of the radius lies 
outside the event horizon on the ``our'' side of the Einstein-Rosen bridge (thus replacing 
the notion of ``classical turning point'' by that of ``radius expectation value''). For the 
decreasing branch the same occurs on the ``other'' side of the Einstein-Rosen bridge. 
It is clear now why the value of total mass-bare mass ratio in quantum case is greater than 
the corresponding classical value. The origin of this is just the replacing of the classical 
turning point by the mean value which is smaller thus giving rise to the increase in the 
total mass.

Following such a line of reasoning we should conjecture that the maximal possible value 
of a total mass $m$ for charge $e$ and principal quantum number $n$ fixed corresponds to 
the situation when the mean value of the radius of the shell lies at the event horizon 
making its collapse possible. The further increase in the bare mass $m$ will lead to the 
wormhole with smaller total mass.

Generalizing, we can introduce the notion of the quantum black hole states.
For given values of electrical charge $e$ and principal quantum number $n$ the quantum 
black hole state is the state with maximal possible total mass $m$.
Thus, for the quantum black hole states we have, instead of inequality (\ref{branch}) the 
following equation
\begin{equation}
3(\kappa M^2)^2 - 4\kappa M^2e^2 + e^4 - 4n^2 = 0 ,
\label{brancheq}
\end{equation} 
the solution of which subject to the condition $\kappa M^2-e^2>0$ is
\begin{equation}
\kappa M^2 = \frac{2}{3}e^2 + \frac{1}{3}\sqrt{e^4 + 12n^2} .
\label{root}
\end{equation}
For the black hole mass spectrum we get eventually
\begin{eqnarray}
m_{BH} = \frac{1}{\sqrt{\kappa}} \sqrt{\frac{4\sqrt{3}}{9}n \left(1 + \frac{e^4}{12n^2}
\right)^{3/2} + \frac{2}{3}e^2 - \frac{e^6}{54n^2}}\\
n = 1, 2, 3,...\nonumber
\label{BH}
\end{eqnarray}
For the uncharged black hole with $e=0$ we have immediately
\begin{equation}
m = \frac{2}{\stackrel{4}{\sqrt{27}}}\sqrt{n}M_{Pl} .
\label{BHm}
\end{equation}
Two limiting cases are of interest. In the first case
\begin{eqnarray}
\frac{e^4}{12n^2} \ll 1,\nonumber\\
\nonumber\\
m_{BH} = \frac{2}{\stackrel{4}{\sqrt{27}}}\frac{\sqrt{n}}{\sqrt{\kappa}}
\left(1 + \frac{\sqrt{3}}{4}\frac{e^2}{n} - \frac{1}{32}\frac{e^4}{n^2} + \ldots\right) ,
\label{one}
\end{eqnarray}
and the charge of the black hole gives rise only to the small corrections to the mass 
spectrum of the Schwarzschild black hole. The corresponding expressions for the bare mass 
$M$ and the ratio $m/M$ are, respectively,
\begin{eqnarray}
M = \stackrel{4}{\sqrt{\frac{4}{3}}}\frac{\sqrt{n}}{\sqrt{\kappa}}
\left(1 + \frac{\sqrt{3}}{6}\frac{e^2}{n} - \frac{1}{48}\frac{e^4}{n^2} + \ldots\right)
\nonumber\\
\nonumber\\
\frac{m}{M} = \sqrt{\frac{2}{3}}
\left(1 + \frac{\sqrt{3}}{12}\frac{e^2}{n} - \frac{5}{96}\frac{e^4}{n^2} + \ldots\right) .
\label{MmM}
\end{eqnarray}
In the second case
\begin{eqnarray}
\frac{e^4}{12n^2} \gg 1,\nonumber\\
\nonumber\\
m_{BH} = \frac{|e|}{\sqrt{\kappa}}\left(1 + \frac{n^2}{2e^4} + \ldots\right) .
\label{two}
\end{eqnarray}
The corresponding expressions for $M$ and $m/M$ are, respectively,
\begin{eqnarray}
M = \frac{|e|}{\sqrt{\kappa}}\left(1 + \frac{n^2}{e^4} + \ldots\right) ,\nonumber\\
\nonumber\\
\frac{m}{M} = 1 - \frac{n^2}{2e^4} + \ldots .
\label{MmM2}
\end{eqnarray}
It is interesting that in this second limiting case the black hole mass takes values nearly 
the mass of a classical extreme Reissner-Nordstrom black hole.
It is easily seen that in both limiting cases $|\bigtriangleup m|/m\ll 1$, where 
$|\bigtriangleup m|$ is the difference in masses for nearby energy (mass) levels. So, these 
limits are essentially quasiclassical ones, and the corresponding mass spectra should be 
compatible with the existence of Hawking's radiation. This is indeed the case as will be 
shown in the next section.
\newpage
{\large \bf XII. Quantum Black Holes and Hawking's Radiation.}
\vskip1cm
In 1972 J.Bekenstein \cite {Jacob} recognized that the classical law that the area of a 
black hole horizon can not decrease has the striking resemblance of the second law of 
thermodynamics that the entropy of an isolated system can not decrease. There began the 
thermodynamical era of the black hole physics. The four laws of black hole mechanics has 
been formulated and proved \cite{fourlaws}. 

This era became a quantum era in 1974 after S.W.Hawking \cite{Hawking} discovered that due 
to quantum fluctuations black holes emit a blackbody radiation with the temperature which 
only by a numerical factor differs from Bekenstein's temperature, and the corresponding 
black hole entropy is just one fourth of the event horizon area. 
The starting point of our further investigation is the first law of the black hole 
thermodynamics which reads (for a spherically symmetric charged black hole) 
\begin{equation}
dm = \frac{1}{4} \Theta dA + \Phi de ,
\label{dm}
\end{equation} 
where $dm, dA, de$ are changes in a total mass $m$, event horizon dimensionless area $A$ 
and electric charge $e$ of the black hole, respectively, $\Theta$ is the temperature, 
$\Phi$ is a Coulomb potential at the horizon, and 
\begin{equation}
A = \frac{4\pi}{\kappa} r_+^2 ,
\label{A}
\end{equation} 
\begin{equation}
\Phi = \frac{4\pi er_+}{A} ,
\label{-phi}
\end{equation} 
\begin{equation}
\Theta = \frac{2}{\kappa A}(r_+ - \kappa m) ,
\label{theta}
\end{equation} 
\begin{equation}
r_+ = \kappa m + \sqrt{\kappa^2m^2 - \kappa e^2} .
\label{r+}
\end{equation} 
The temperature of the Reissner-Nordstrom black hole is low in two different cases: when 
the mass of the black hole goes to infinity (for fixed charge) and when it goes to the 
minimal possible value $\sqrt{\kappa}m=e$ (extreme black hole). In both cases the back 
reaction of the radiation emitted is negligible, so the approximation of the static purely 
Reissner-Nordstrom black hole is valid. Moreover, it is just the quasiclassical limit of 
the (future) complete quantum picture in the sense that $(|\bigtriangleup m|/m\ll 1)$.
The subsequent line of reasoning is after V.Mukhanov \cite{Mukhanov}. Here we extend his 
arguments to include the charge black holes, in particular, the nearly extreme 
Reissner-Nordstrom black holes.

Let us suppose that the black holes are, in fact, quantizes, so their total mass and, hence, 
the area of the horizon are function of an integer number $n$. And let the black hole 
emits the quantum energy $\omega$ due to transition from the $n$-th state to the $(n-1)$-th 
state. The ``typical'' energy of this quantum minus the work done against the Coulomb 
attraction should be proportional to the temperature, i.e.
\begin{equation}
\omega_{n,n-1} - \Phi de = \pi\alpha\Theta ,
\label{omega}
\end{equation} 
where $de$ is the charge of the quantum in question, and $\alpha$ is a slowly varying 
function of the integer quantum number $n$. Inserting, then, Eqn.(\ref{omega}) into 
Eqn.(\ref{dm}) we have $(\omega_{n,n-1}=dm)$
\begin{equation}
dA = 4\pi\alpha ,
\end{equation}
or
\begin{equation}
dA = 4\pi\alpha dn ,
\label{dA}
\end{equation}
Integrating we get
\begin{equation}
A = 4\pi\tilde\alpha n + 4\pi C ,
\label{area}
\end{equation}
where $C$ is a constant of integration, and $\tilde\alpha$ is some another slowly varying 
function of $n$. We consider two limits, $n\gg C$ and $n\ll C$ . For the first case we 
assume $\tilde\alpha$ has the following expansion
\begin{eqnarray}
\tilde\alpha = \alpha_1 + \beta_1\frac{C}{n} + \gamma_1\frac{C^2}{n^2} =\ldots ,\\
\nonumber\\
\frac{C}{n} \ll 1 ,\nonumber
\label{ll}
\end{eqnarray}
and for the latter one
\begin{eqnarray}
\tilde\alpha = \alpha_2 + \beta_2\frac{n}{C} + \gamma_2\frac{n^2}{C^2} =\ldots ,\\
\nonumber\\
\frac{C}{n} \gg 1 .\nonumber
\label{gg}
\end{eqnarray}
Let us write the expression for a black hole mass $m$ in terms of a horizon area $A$ 
making use of Eqns.(\ref{A}) and (\ref{r+}), 
\begin{equation}
m = \frac{1}{\sqrt{\kappa}} \frac{\frac{A}{4\pi} + e^2}{2\sqrt{\frac{A}{4\pi}}} .
\label{mA}
\end{equation}
Inserting then Eqn.(\ref{area}) for $A$, using expansions (\ref{ll}) and (\ref{gg}) for 
$\tilde\alpha$ we get
\begin{eqnarray}
m &=& \frac{\sqrt{n}}{2\sqrt{\kappa}}\sqrt{\alpha_1} \left(1 + 
\left(\frac{\beta_1+1}{2\alpha_1} + \frac{e^2}{C\alpha_1}\right) \frac{C}{n} \right.
\nonumber \\
&+& 
\left. \left(\frac{\gamma_1}{2\alpha_1} - \frac{(\beta_1+1)^2}{8\alpha_1^2} - 
\frac{\beta_1+1}{2\alpha_1^2}\frac{e^2}{C}\right)\frac{C^2}{n^2} +\ldots\right) ,\\
\nonumber\\
&&n \gg C\nonumber
\label{mll}
\end{eqnarray}
and
\begin{eqnarray}
m &=& \frac{\sqrt{C}}{2\sqrt{\kappa}} \left(1 + \frac{e^2}{C} + 
\frac{\alpha_2}{2}\left(1 - \frac{e^2}{C}\right) \frac{n}{C} \right. \nonumber \\
&+& \left.
\left(\frac{\beta_2}{2}\left(1 - \frac{e^2}{C^2}\right) - \frac{1}{8}\alpha_2^2
\left(1 - \frac{3e^2}{C^2}\right)\right)\frac{n^2}{C^2} +\ldots\right) ,\\
\nonumber\\
&&n \ll C\nonumber
\label{mgg}
\end{eqnarray}
Comparing this with the expansion obtained from the black hole spectrum, we have
\begin{eqnarray}
C &=& e^2\nonumber\\
\alpha_1 = \frac{16}{3\sqrt{3}},\qquad \beta_1 &=& -\frac{1}{3},\qquad \gamma_1 = 
\frac{5\sqrt{3}}{144},...\nonumber\\
\alpha_2 = 2,\qquad \beta_2 &=& 2,\qquad \gamma_2 = -15,... 
\end{eqnarray}
Comparing values for $\alpha_1$ and $\alpha_2$ we see that $\tilde\alpha$ is indeed a 
slowly varying function.

Thus, we see that in the limiting case of low temperature our black hole spectrum is 
compatible with the existence of Hawking's radiation.
Finally, we would like to note that our black hole spectrum obeys also the third law of 
thermodynamics. Indeed, for a nearly extreme Reissner-Nordstrom black hole the lowest 
energy level (for $n=1$) always exceeds the critical value $(=|e|/\sqrt{\kappa})$, so 
the zero temperature state is not accessible.
\newpage
{\large\bf Discussion.}
\vskip1cm
In the concluding remarks we would like to discuss shortly the obtained results.

1.The appearance of the Schroedinger equation in finite differences is interesting  by 
itself. And the fact that such an equation with Coulomb potential appeared to be exactly 
solvable seems miraculous. We believe that this is not accidental and there is some some 
symmetries that made the solution possible. Anyway, this point needs further investigation. 

2.The mass spectrum for the self-gravitating thin shells can be obtained from Dirac's 
spectrum for hydrogen atom by putting zero the so called radial quantum number (e.i. for 
the critical value of the total angular momentum).

3.To extract the black hole spectrum from the thin shell spectrum we introduced the 
definition of the quantum black holes as a marginal case between quantum thin shells 
which do not collapse and wormholes. In quasiclassical regime this definition can be 
interpreted in that way that the expectation value of the radius of the shell is equal 
to the Schwartzschild radius. It is interesting that the special role played by the 
shells ``lying'' at the black hole horizon is also manifested in the processes of vacuum 
phase transitions which result in a black hole creation. In paper \cite{O3} it was shown 
in a quasiclassical approximation that the O(3)-symmetrical vacuum phase transition with 
a black hole initially present or a blaÓk hole formation from ``nothing'' may take place 
only if the new vacuum bubble shell is created exactly at the event horizon.

4.We would like to emphasize once more the similarities an differences of our picture from 
the hydrogen atom models. In classical physics the electron in a hydrogen atom collapses 
radiating away electromagnetic energy. In the quantum mechanical picture it is stable 
unless something will force it to emit a photon and change its orbit, and there exists 
the lowest possible energy level which is by definition stable. Our self-gravitating shell 
collapses without any radiation in classical physics. The quantum mechanical picture is 
analogous to that of hydrogen atom. There exists also the lowest possible energy level. 
But unlike the hydrogen atom spectrum, our spectrum depends not only on the charge 
(which is already quantized for the electron) but also on a continuous parameter $M$, 
the bare mass of the shell. Moreover, we know that in the quasiclassical limit the black 
hole emits particles, thus acting as a source of a blackbody radiation. This radiation 
is caused by the vacuum fluctuations of different fields and by existence of the event 
(apparent) horizon. It seems that the radiation is an inherent feature of the 
gravitational collapse. And to solve the problem of gravitational collapse we need to take 
into account the vacuum fluctuations and spontaneous emission caused by them. This will 
require the self-consistent solution of the gravitational and field theoretical (or string 
theory) problems.

5.We found that each eigenvalue of a total mass (energy) is infinitely degenerate. Clearly, 
the origin of such a degeneracy lies in freezing all but one dynamical degrees of 
freedom. Our experience in quantum mechanics suggests that after restoring the frozen 
degrees of freedom this degeneracy disappears. As a result we expect that the ``true'' 
spectrum for thin shell masses will depend on more than one quantum number. Due to infinite 
degeneracy of the ``frozen'' spectrum the ``true'' one will, in fact, be quasicontinuous. 
And, as a consequence, the Hawking's evaporation spectrum will be quasicontinuous as well. 
The existence of the ground state with nonzero mass means that the evaporation process 
should stop when the mass of a black hole approaches the minimum possible value 
$(\approx 0.9 M_{Pl})$ . And we may conjecture that the final state of the black hole 
evaporation will be characterized by the minimal value of mass but different wave 
functions of a ground state depending on details of mass distribution and evaporation 
process. We may call these wave functions quantum black hole hairs. This unexpected feature 
of the black hole mass spectrum (its infinite degeneracy) reminds the problem of so called 
hidden parameters in quantum mechanics and may be helpful in resolving the well known 
information paradox in a black hole physics.

6.The very fact that our formulas may be obtained in the quasiclassical limits from the 
very general thermodynamical grounds encourages us to believe that their range of validity 
is not confined to the specific model used for derivation.
7.There exists the minimal possible mass for the black hole. For the uncharged case it is 
slightly less that the Planckian mass. Such a minimal black hole mass was predicted by 
M.A.Markov \cite{Markov}. It is also the same for small electric charges and tends to the 
extreme value for charges comparable (in suitable units) to the black hole mass. For mass 
values less that this minimum there are no black hole states at all, and for all known 
elementary particles this is the case. It may appear that from the point of view of the 
spacetime structure all of them are on the ``other'' side of the Einstein-Rosen bridge, 
forming wormholes.  
\vskip1cm
{\large \bf Acknowledgments.}
\vskip1cm
I am grateful to I.Ya.Aref'eva,~ A.O.Barvinsky,~ T.Damour,~
 B.DeWitt, G.Esposito-Farese,~ 
V.P.Frolov,~ D.Yu.Grigor'ev, ~I.M.Khalatnikov,~ Rocky Kolb,~~ V.A.Kuzmin,~~ A.N.Kuznetzov, 
~~G.B.Pivovarov,~~ V.A.Rubakov,\\ 
D.V.Semikoz,~~~ M.E.Shaposhnikov, ~~~A.A.Starobinsky, 
~~~I.I.Tkachev,\\ 
F.V.Tkachov,~ G.A.Vilkovisky, ~I.V.Volovich for many helpful discussions. Special thanks 
to D.O.Ivanov.
 
\end{document}